\documentclass[twocolumn]{aastex631}

\usepackage{amsfonts,amsmath,amssymb,amsthm}
\usepackage{longtable}
\usepackage{color}
\usepackage[T1]{fontenc}
\usepackage{ae,aecompl}
\usepackage{graphicx}	
\usepackage{mathrsfs}
\usepackage{tabularx}
\usepackage{dcolumn}
\usepackage{bm}
\usepackage{epsfig}
\usepackage{float}
\usepackage{varwidth}
\usepackage{soul}
\usepackage{ulem}
\usepackage{cancel}
\usepackage{mathtools}
\usepackage{hyperref}

\newcommand{\vect}[1]{{\boldsymbol{#1}}}

\renewcommand{\d} {\mathrm{d}}

\newcommand{\bnabla}{\boldsymbol{\nabla}}

\shortauthors{Franci et al.}

\begin{document}

\title{Anisotropic electron heating in turbulence-driven magnetic reconnection in the near-Sun solar wind}

\author[0000-0002-7419-0527]{Luca Franci}
\affiliation{School of Physical and Chemical Sciences, Queen Mary University of London, London, UK}
\affiliation{National Institute for Astrophysics (INAF) - Institute for Space Astrophysics and Planetology (IAPS), Rome, Italy}
\author[0000-0002-7969-7415]{Emanuele Papini}
\affiliation{National Institute for Astrophysics (INAF) - Institute for Space Astrophysics and Planetology (IAPS), Rome, Italy}
\author[0000-0001-9293-174X]{Alfredo Micera}
\affiliation{Solar-Terrestrial Centre of Excellence–SIDC, Royal Observatory of Belgium, Brussels, Belgium}
\author[0000-0002-3123-4024]{Giovanni Lapenta}
\affiliation{Centre for Mathematical Plasma Astrophysics, KU Leuven, Leuven, Belgium}
\author[0000-0002-5608-0834]{Petr Hellinger}
\affiliation{Astronomical Institute, Czech Academy of Sciences, Prague, Czech Republic}
\affiliation{Institute of Atmospheric Physics, Czech Academy of Sciences, Prague, Czech Republic}
\author[0000-0001-7489-8843]{Daniele Del Sarto}
\affiliation{Institut Jean Lamour, UMR 7198 CNRS – Universit\'e de Lorraine, Nancy, France}
\author[0000-0002-8175-9056]{David Burgess}
\affiliation{School of Physical and Chemical Sciences, Queen Mary University of London, London, UK.}
\author[0000-0002-1322-8712]{Simone Landi}
\affiliation{Dipartimento di Fisica e Astronomia, Universit\`a degli Studi di Firenze, Sesto Fiorentino, Italy}

\begin{abstract}
We perform a high-resolution two-dimensional fully-kinetic numerical simulation of a turbulent plasma system with observation-driven conditions, in order to investigate the interplay between turbulence, magnetic reconnection, and particle heating from ion to sub-electron scales in the near-Sun solar wind. We find that the power spectra of the turbulent plasma and electromagnetic fluctuations show multiple power-law intervals down to scales smaller than the electron gyroradius. Magnetic reconnection is observed to occur in correspondence of current sheets with a thickness of the order of the electron inertial length, which form and shrink due to interacting ion-scale vortexes. In some cases, both ion and electron outflows are observed (the classic reconnection scenario), while in others ---typically for the shortest current sheets--- only electron jets are presents (``electron-only reconnection''). At the onset of reconnection, the electron temperature starts to increase and a strong parallel temperature anisotropy develops. This suggests that in strong turbulence electron-scale coherent structures may play a significant role for electron heating, as  impulsive and localized phenomena such as magnetic reconnection may transfer energy from the electromagnetic fields to particles more efficiently than damping mechanisms related to interactions with wave-like fluctuations.
\end{abstract}

\keywords{Space plasmas --- Plasma astrophysics --- Solar wind --- Interplanetary turbulence }

\date{\today}

\section{Introduction}
\label{sec:introduction}
The solar wind is a highly magnetized and nearly collisionless plasma flow that originates in the Sun's corona \citep{Marsch_2006} and blows in our solar system, filling the the whole heliosphere. While expanding away from the Sun, its temperature gradually decreases with heliocentric distance. Such temperature decrease, however, is smaller than what is expected in the case of a purely adiabatic expansion \citep[e.g.,][and therefernces therein]{Verscharen_al_2019}.
This represents the observational evidence that local heating and particle energisation mechanisms are at play in the solar wind \citep{Goldstein_al_2015}. 
Despite decades-long investigation by different heliospheric missions, understanding the origin  of solar wind heating is one of the long-standing open issues in space plasma physics. Among others, turbulent dissipation is the most promising candidate to explain the observed heating \citep[e.g.,][]{Kiyani_al_2015,Parashar_al_2015}.

Turbulence is routinely observed in the solar wind \citep[e.g.,][]{Bruno_Carbone_2013}. In-situ spacecraft observations return power spectra of magnetic fluctuations exhibiting a power-law behavior over many decades in frequency \citep[e.g.,][]{Tu_Marsch_1995,Bruno_Carbone_2013,Chen_al_2013b,Kiyani_al_2015}. This is the expression of a turbulent energy cascade:
the magnetic and kinetic energy stored at the largest scales is transferred (i.e., "cascades") via nonlinear interactions to progressively  smaller and smaller scales, down to particles characteristic scales. As the solar wind plasma is collisionless, the energy cascade must terminate via mechanisms that are alternative to particle collisions.

Typically, the power spectrum of the turbulent magnetic fluctuations has a spectral index (i.e., a slope), $\alpha$, that decreases with increasing frequency.
At large scales/low frequencies (the so-called ``inertial range''), where the plasma can be described as a magnetized fluid within the framework of magnetohydrodynamics (MHD), both in-situ observations and numerical simulations report $\alpha \in [-5/3,-3/2]$, depending on the level of imbalance between outward and inward propagating Alfv\'en waves and on the heliocentric distance \citep[e.g.,][]{Boldyrev_al_2011,Chen_al_2013,Chen_2016,Chen_al_2020}. 
As we approach particles characteristic scales, where kinetic effects become important, the picture gets increasingly complex. In the range across and just below the ion characteristic scales (known as ``spectral break'' or ``transition region''), both observations and  simulations recover $\alpha \in [-4,-2]$, depending on plasma parameters such as the ion plasma beta and the level of turbulent fluctuations, $\alpha \simeq -2.8$ being the value more frequently observed \citep[e.g.,][]{Bruno_al_2014,Franci_al_2016b,Verscharen_al_2019}. When the turbulent cascade reaches the electron scales, $\alpha$ further decreases \citep[e.g.,][]{Alexandrova_al_2009,Sahraoui_al_2013}. 
At present, there is no consensus on the spectral properties of the magnetic fluctuations around and below the electron characteristic scales and even the description of their power spectrum in terms of a spectral index is controversial. In particular, two alternative phenomenological descriptions have been presented, which model the power spectrum of the magnetic field either as a power law with an exponential cut-off \citep{Alexandrova_al_2009,Alexandrova_al_2012,Alexandrova_al_2021}, or as two power laws with different slopes at scales larger and smaller than the electron gyroradius \citep{Sahraoui_al_2013,Huang_Sahraoui_2019}.
While the spectral index observed in the inertial range can be explained in terms of the Kolmogorov phenomenology \citep{Kolmogorov_1941}, eventually extended to include MHD effects \citep{Iroshnikov_1963,Kraichnan_1965,Goldreich_Sridhar_1995}, at sub-ion scales we still lack a definite explanation.

In the framework of wave-like turbulence, the energy cascade at kinetic scales is mediated by fluctuations whose properties resemble those of either kinetic Alfv\'en waves of whistlers waves \citep{Howes_al_2008a,Schekochihin_al_2009}. The mechanisms responsible for dissipating energy and terminating the cascade, thus leading to particle heating, can be either resonant, as for Landau and cyclotron damping \citep[e.g.,][]{He_al_2015,Sulem_al_2016,Howes_al_2018,Chen_2019}, or non-resonant, as in the case of stochastic heating\citep[e.g.,][]{Chandran_al_2013,Vech_al_2017,Martinovic_al_2020}. 
Recent numerical studies \citep[][]{Parashar_al_2010,Papini_al_2021} have revealed, however, that the turbulent structures at sub-ion scales are characterized by very-low temporal-frequency features, not directly connected to wave activity. This strongly supports the idea that coherent structures such as vortexes, discontinuities, and thin intense current sheets, can play a key role in mediating the turbulent cascade \citep[e.g.,][and references therein]{Gosling_2007,Greco_al_2012,Karimabadi_al_2013,Osman_al_2014,Franci_al_2016b,Cerri_Califano_2016,Perrone_al_2016,Wan_al_2016,Franci_al_2017,Yang_al_2017,Camporeale_al_2018,Papini_al_2020,Agudelo_al_2021,Papini_al_2021,Charles_Vasquez_2021}.
Particle energization in collisionless plasmas can proceed via the pressure-strain
coupling, which starts to play an important role around the ion scales and by which the kinetic energy of a spatially inhomogeneous flow is anisotropically transferred to internal energy \citep{DelSarto_al_2016,Yang_al_2017,DelSarto_Pegoraro_2018,Matthaeus_al_2020,Bandyopadhyay_al_2021,Hellinger_al_2022}.

Another promising mechanism for energy dissipation and particle heating in turbulent plasmas is magnetic reconnection \citep[e.g.,][]{Retino_al_2007,Gosling_2007,Ergun_al_2017,Voros_al_2017,Eastwood_al_2018,Stawarz_al_2019,Zhou_al_2021}. Through reconnection, the energy contained in the magnetic field is released into the plasma, triggered by a topological change in the magnetic field configuration and the subsequent relaxation of the magnetic field lines. Reconnection is usually accompanied by the appearance of ion and electron outflow jets. Recently, however, in-situ observations by Magnetospheric Multiscale (MMS) in the Earth's magnetosheath have revealed some peculiar magnetic reconnection events producing electron jets with no ion counterpart (``electron-only reconnection'') \citep{Phan_al_2018}. These were observed in correspondence of thin electron-scale current sheets embedded in strong turbulent fluctuations. Further observational and numerical studies have shown that this is likely to occur when the correlation length of the turbulence is of the order of just a few ion inertial lengths, thus current sheets tend to form and reconnect at scales at which the ions have already decoupled from the magnetic field \citep{Stawarz_al_2019,Califano_al_2020,SharmaPyakurel_al_2019,Stawarz_al_2022}.

As reconnection converts magnetic energy into particle kinetic and thermal energy, it likely plays a major role in the evolution and dissipation of turbulence  \citep[][]{Servidio_al_2009,Matthaeus_Velli_2011}. Indeed, turbulence and reconnection are observed to be intimately linked to each other. Numerical simulations also show that, on the one hand, turbulence develops in reconnection outflows \citep{Karimabadi_al_2013,Pucci_al_2017,Lapenta_al_2020} and the turbulent cascade at sub-ion scales can be directly triggered by reconnection events \citep{Franci_al_2017}. On the other hand, the interaction between turbulent structures spontaneously generates and squeezes thin intense current sheets, until they eventually reconnect \citep[e.g.,][]{Matthaeus_1986,Biskamp_Bremer_1994,Wei_al_2000,Gosling_2007,Servidio_al_2009,Servidio_al_2011,Franci_al_2017,Cerri_Califano_2016,Papini_al_2019,Stawarz_al_2019,Agudelo_al_2021}.
Magnetic reconnection is also linked to particle temperature anisotropy, as this is expected to be correlated to the hyperbolic flow close to a reconnecting X-point \citep{Cai_Lee_1997,Brackbill_2011,DelSarto_al_2016}. Anisotropic electron heating and non-gyrotropic velocity distribution functions have indeed been observed both in numerical simulations of single reconnection events \citep{Cai_Lee_1997,Aunai_al_2013,Fulvia_al_2018,Sladkov_al_2021} and in Cluster measurements of a reconnecting thin current sheet in the Earth's magnetotail \citep{Retino_al_2008}. 

Fully-kinetic simulations of collisionless turbulent plasmas, which retain both ion and electron kinetic effects, represent an invaluable tool for investigating the turbulent energy cascade down to electron scales \citep[e.g.,][]{Groselj_al_2018,Gonzalez_al_2019,Roytershteyn_al_2019,Cerri_al_2019}, its interplay with magnetic reconnection \citep[e.g.,][]{Karimabadi_al_2013,Pucci_al_2017,Pucci_al_2018,Adhikari_al_2021,Agudelo_al_2021}, as well as the role of electron-scale coherent structures in dissipating energy and heating particles \citep[e.g.,][]{Camporeale_Burgess_2011,Parashar_al_2015,Yang_al_2017,Arro_al_2021,Bandyopadhyay_al_2021,Yang_al_2022}. They have also provided numerical evidence for an enhancement of the electron parallel temperature anisotropy in the outflows of strong reconnection events, which occurred spontaneously as the result of the interactions between sub-proton scale turbulent structures \citep{Camporeale_Burgess_2011,Haynes_2014}.

In this work, we investigate the development of turbulence, its interplay with magnetic reconnection, and particle heating by means of a high-resolution 2D fully-kinetic simulation of strong plasma turbulence in the near-Sun solar wind, carried out using the iPic3D code \citep{Markidis_al_2010}. The initial conditions model the average plasma environment encountered by the Parker Solar Probe (PSP) spacecraft during its first solar encounter at about 36 solar radii from the Sun \citep{Bale_al_2019}. The plasma system size is 32 times the ion inertial length, which allows us to accurately model the turbulent energy cascade from ion scales down to sub-electron scales, providing predictions for the spectral properties of the turbulent fluctuations at scales that cannot be resolved by PSP. To overcome the current unfeasibility of resolving the electron scales while concurrently modelling the large MHD scales, we compare the turbulent spectral properties  of the fully-kinetic simulation with those obtained from a hybrid-kinetic simulation \citep[performed with the CAMELIA code][]{Franci_al_2018b}, which models a much larger system with a lower resolution, but employing the same physical plasma conditions.


The paper is organized as follows. In Sec.\ref{sec:setup}, we describe the numerical dataset, providing details on the simulation setup and fundamental plasma parameters. In Sec.\ref{sec:results}, we present our results, in terms of spectral properties of the turbulent energy cascahde, particle heating, and the occurrence of standard and electron-only reconnection events. Finally, we discuss our findings and draw our conclusions in Sec.\ref{sec:conclusions}.

\section{Numerical dataset}
\label{sec:setup}

We performed a 2D particle-in-cell (PIC) simulation of plasma turbulence using the semi-implicit code iPIC3D~\citep{Markidis_al_2010}, a fully kinetic code which solves the Vlasov-Maxwell equations for a non-relativistic plasma of ions and electrons. Ipic3D employs an implicit scheme for the temporal integration of the Vlasov–Maxwell system~\citep{Brackbill1982} which removes the numerical stability constraints typical of explicit schemes~\citep{Cohen1981, Hockney1988}, so that it is possible to retain small spatiotemporal scales in an approximate way without the need to resolve the Debye length and to include the speed of light in the Courant condition for the temporal integration. This allows us to retain all kinetic effects, which are vital to describe correctly the overall evolution of the system, while still employing a simulation box whose size is an order of magnitude larger than the ion characteristic scales. 

The initial condition consists of a uniform plasma composed of electrons and ions (assumed to be only protons, embedded in and ambient magnetic field $\vect{B}_0$.  

We use the following normalization units: the magnitude of the ambient field $B_0 = |\vect{B}_0|$ for the magnetic fluctuations, the initial plasma density $n_0 = n_{i,0} = n_{e,0} $ for the density fluctuations, the (ion) Alfv\'en speed $v_A = d_i \Omega_i = B_0/ \sqrt{4 \pi n_0 m_i}$ for the velocity fluctuations, the inverse of the proton plasma frequency $\Omega_i = e B_0/(m_i c)$ for time, the proton inertial length $d_i = v_A/\Omega_i$ for lengths. The plasma beta for a given plasma species is
$\beta_{i,e} = 8 \pi n_{i,e} K_B T_{i,e} / B_0^2$. 
The electron spatial and temporal characteristic scales are related to the ion ones through the proton-to-electron mass ratio $\mu = m_i/m_e$ as $d_e = d_i/\sqrt{\mu}$ and $\Omega_e = \mu \Omega_i$. Quantities and symbols used in these definitions are: the speed of light $c$, the ion and electron number densities $n_{i,e}$, the magnitude of the electronic charge $e$, the proton and electron masses $m_{i,e}$, the Boltzmann's constant $K_B$, and the proton and electron temperatures $T_{i,e}$.

The 2D computational domain consists of $2048^2$ grid points with a spatial resolution $\Delta {x,y} =
d_i/64$ and a size
$L_{x,y} = 32 \, d_i$. The time step for the particle advance is $\Delta t = 0.000625 \, \Omega_i^{-1} = 0.0625 \, \Omega_e^{-1}$. We employ a reduced ion-to-proton mass ratio of $\mu = m_i /m_e = 100$ in order to decrease the computational cost of the simulation and $1024$ ions and $8192$ electrons per cell to reduce the noise level at small scales. The ambient magnetic field $\vect{B}_0$ is along the $z$-direction. We set its magnitude such that $c/v_{A} = \omega_i/ \Omega_i = 200$ and $c/v_{Ae} = \omega_e/ \Omega_e = 20$ (we recall here that the electron Alfvén speed is ${v}_{{Ae}}={B}_{0}/\sqrt{4\pi {n}_{e}{m}_{e}}$).
In this work, each field $\Psi$ will be decomposed in its perpendicular
(in-plane) component, $\Psi_\perp$, and its parallel (out-of-plane,
along $\vect{z}$) component, $\Psi_{\parallel}$, with respect to the orientation of $\vect{B}_0$. The only exceptions will be the particle temperatures, for which $\perp$ and $\parallel$ will denote directions
with respect to the \textit{local} magnetic field.

Initially, we assume a uniform number density $n = 1$, an ion plasma beta
$\beta_{i} = 0.2$, an electron plasma beta $\beta_{e} = 0.5$, and no ion nor electron temperature anisotropy, i.e. $A_{i,e} =
T_{(i,e)\perp} / T_{(i,e)\parallel} = 1$.  We add an initial spectrum of in-plane Alfv\'enic-like magnetic and ion bulk velocity
fluctuations, composed of modes with wavenumbers in the range $-0.8 \lesssim
k_{x,y} \, d_i \lesssim 0.8$ and random phases. Their initial global amplitude can be estimated in terms of their root-mean-square value as $\delta \vect{B}^{\textrm{rms}} \simeq 0.39 \,B_0$ and $\delta \vect{u}_i^{\textrm{rms}} \simeq 0.34 \, v_A$, so that there is a small initial residual energy (excess of magnetic over ion kinetic energy). Initial electron bulk velocity fluctuations with both perpendicular and parallel components are also imposed, such that $n_0 (\delta \vect{u}_{i,\parallel} - \delta \vect{u}_{e,\parallel}) = - n_0 \delta \vect{u}_{e,\parallel} = \vect{J}_{\parallel} = \bnabla \times \delta \vect{B}_\perp$ and $n_0 (\delta \vect{u}_{i,\perp} - \delta \vect{u}_{e,\perp}) = \vect{J}_{\perp} = \bnabla_\perp \times \delta \vect{B}_\parallel = 0 $. As a consequence, at t = 0 we have $\delta \vect{u}_{e,\perp} = \delta \vect{u}_{i,\perp}$, so that $\delta \vect{u}_{e,\perp}^{\textrm{rms}} \simeq 0.34 \, v_A$.

\begin{figure}
\centering
\includegraphics[width=0.4\textwidth]{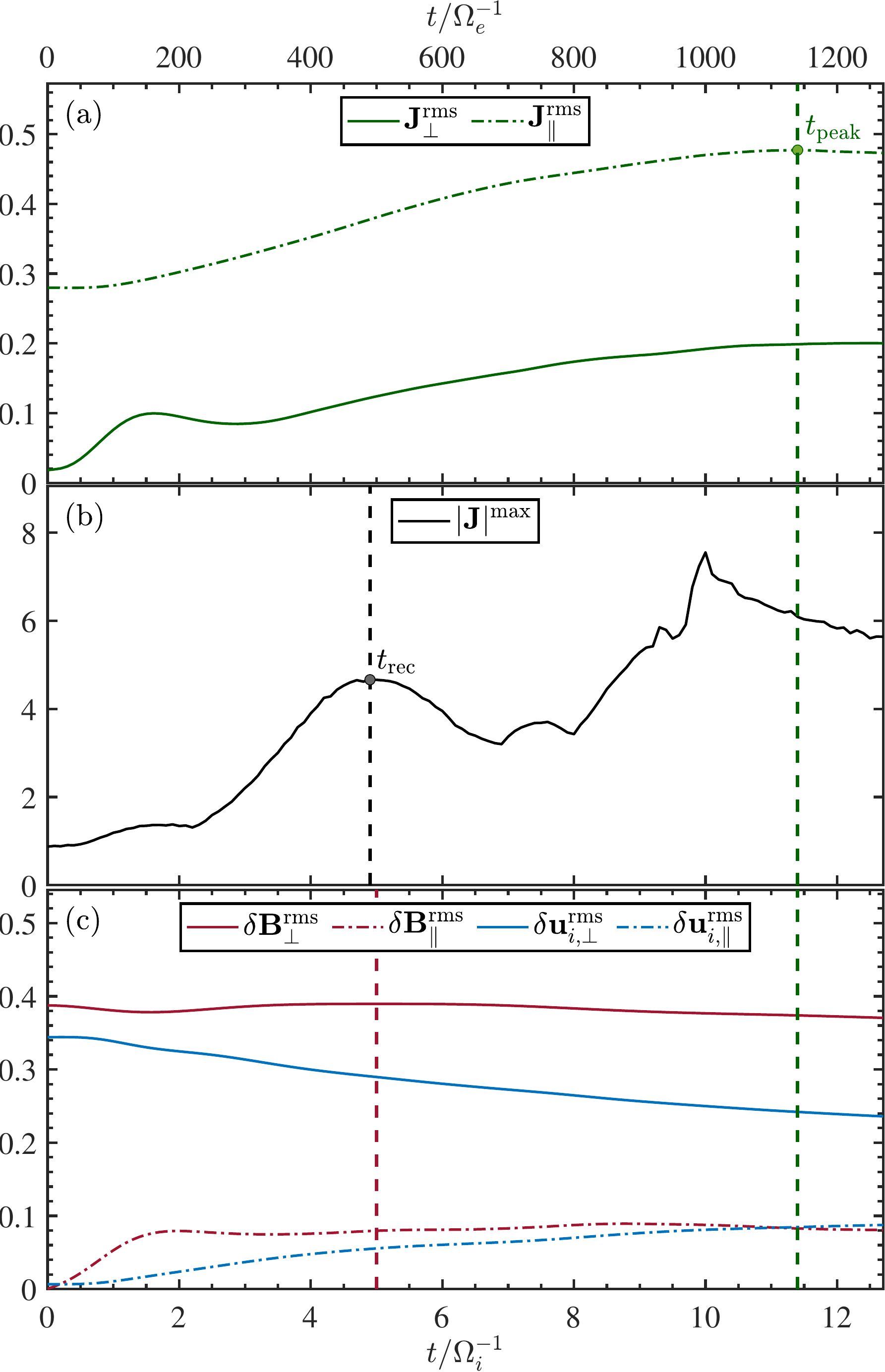}
\caption{Time evolution of some global quantities. \textit{Panel (a)}: root-mean-square (rms) value of the out-of-plane and in-plane components of the current density, $\vect{J}_\parallel$ and $\vect{J}_{\perp}$. \textit{Panel (b)}: maximum of the magnitude of the current density, $|\vect{J}|$, in the whole simulation box. \textit{Panel (c)}: rms of the components of the magnetic fluctuations, $\vect{B}_\perp$ and $\vect{B}_\parallel$, and of ion bulk velocity, $\vect{u}_{i,\perp}$ and $\vect{u}_{i,\parallel}$. In all panels, a vertical dashed green line marks the time when the rms of $\vect{J}_\parallel$ reaches a maximum, i.e., when turbulence has fully developed. In panel (b), a vertical dashed black line marks the time when reconnection events start occurring, which also corresponds to the time when $\vect{B}_\perp$ starts decreasing, indicated by the vertical dashed red line in panel (c).}
\label{fig:time_evolutions}
\end{figure}

\section{Results}
\label{sec:results}

\subsection{Fully developed turbulence at sub-ion scales}
\label{subsec:turbulence}

We let the plasma system evolve from its initial condition until a turbulent energy cascade has fully developed and starts to slowly decay. This occurs when the rms value of the current density, $\vect{J}$, reaches a maximum/plateau~\citep[e.g.][]{Franci_al_2015b,Franci_al_2017}. Figure~\ref{fig:time_evolutions} shows the time evolution of some fundamental space-averaged quantities that allow us to follow the development of the turbulent cascade, identifying different phases.
Figure~\ref{fig:time_evolutions}(a) shows the rms of the components of $\vect{J}$ both parallel and perpendicular to the ambient magnetic field $\vect{B}_0$. The former provides the dominant contribution to the total current and reaches a maximum at $t_{\mathrm{peak}} = 11.4 \, \Omega_i^{-1}$, marked by a green dashed vertical line. The following analysis will therefore be performed at this time. The other component, $\vect{J}_\perp^{\mathrm{rms}}$, is initially almost zero and quickly grows, yet remaining quite smaller than its parallel counterpart at all times. This is related to the rapid development of parallel magnetic fluctuations, as shown below. The eddy turnover time corresponding to the initial injection scale, $k^{\textrm{inj}} d_i \simeq ( k_{x}^2 + k_{y}^2 ) \, d_i \simeq 1.1$ is the time associated to nonlinear energy transfers at $t=0$ and can be estimated as $t^{\textrm{inj}}_{\textrm{NL}} \sim [ k^{\textrm{inj}} \delta B_ k^{\textrm{inj}}]^{-1} \sim [ k^{\textrm{inj}} \delta B^{\textrm{rms}}]^{-1} \sim 2.3 \,\Omega_i^{-1}$. A turbulent energy cascade develops fully from the injection scale down to the electron scales in a few nonlinear times, as $t_{\mathrm{peak}} \sim 5 \, t_{\textrm{NL}}$. This is similar to what previously observed in hybrid simulations with similar initial conditions \citep[e.g.][]{Franci_al_2015a,Franci_al_2015b}, where about 10 nonlinear times were required. 

\begin{figure*}
\centering 
\includegraphics[width=0.48\textwidth]{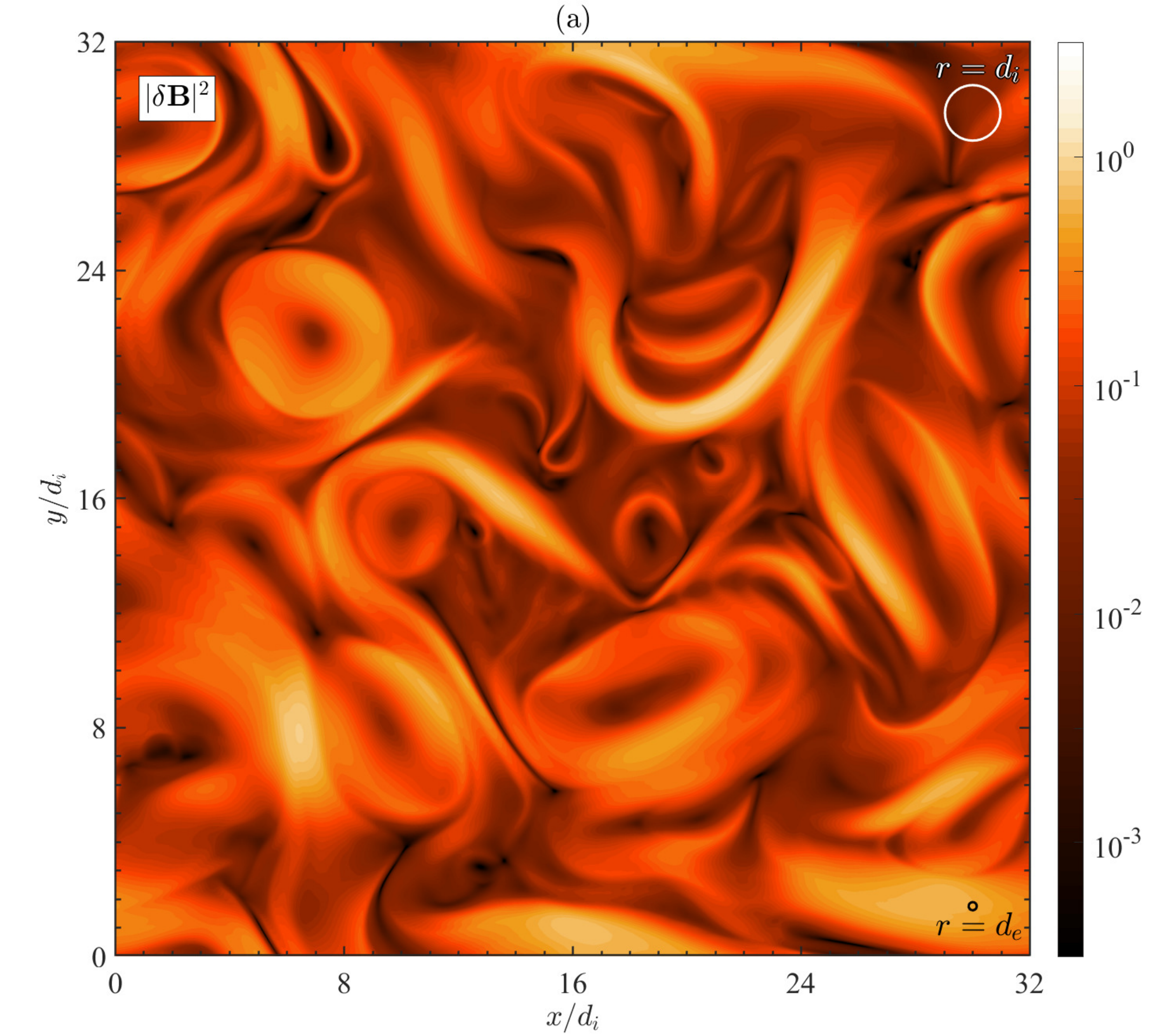}
\includegraphics[width=0.48\textwidth]{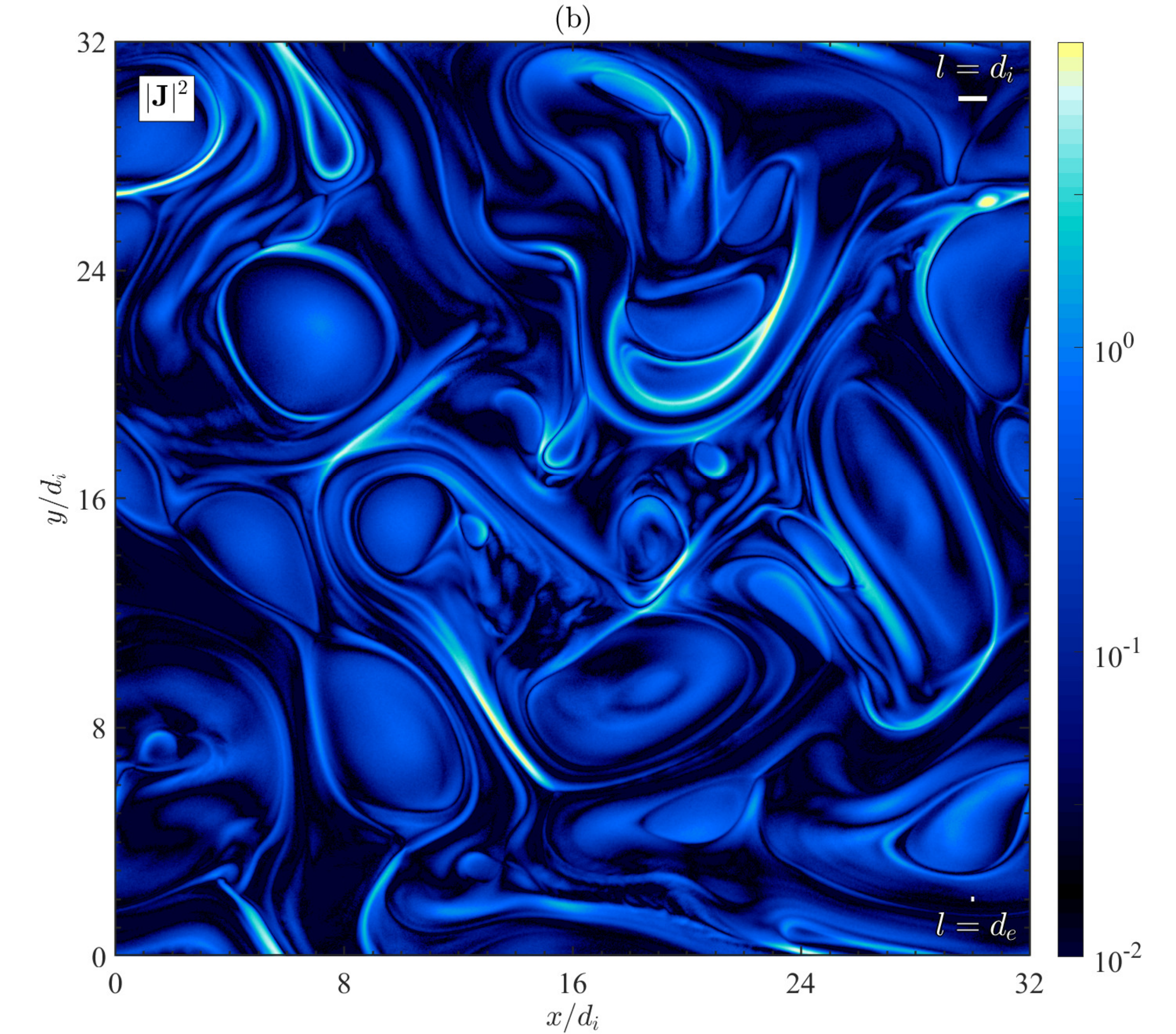}\\
\includegraphics[width=0.48\textwidth]{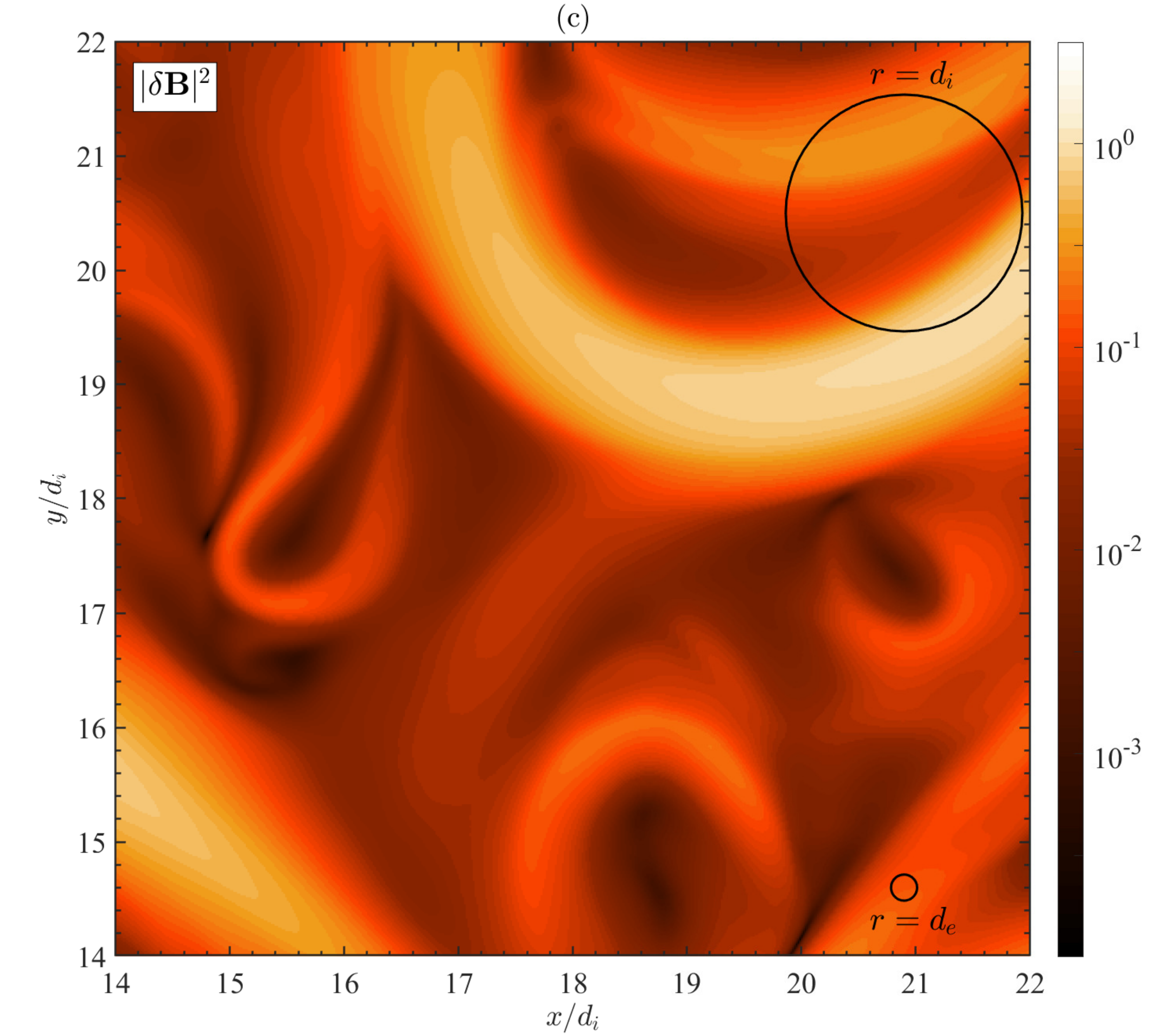}
\includegraphics[width=0.48\textwidth]{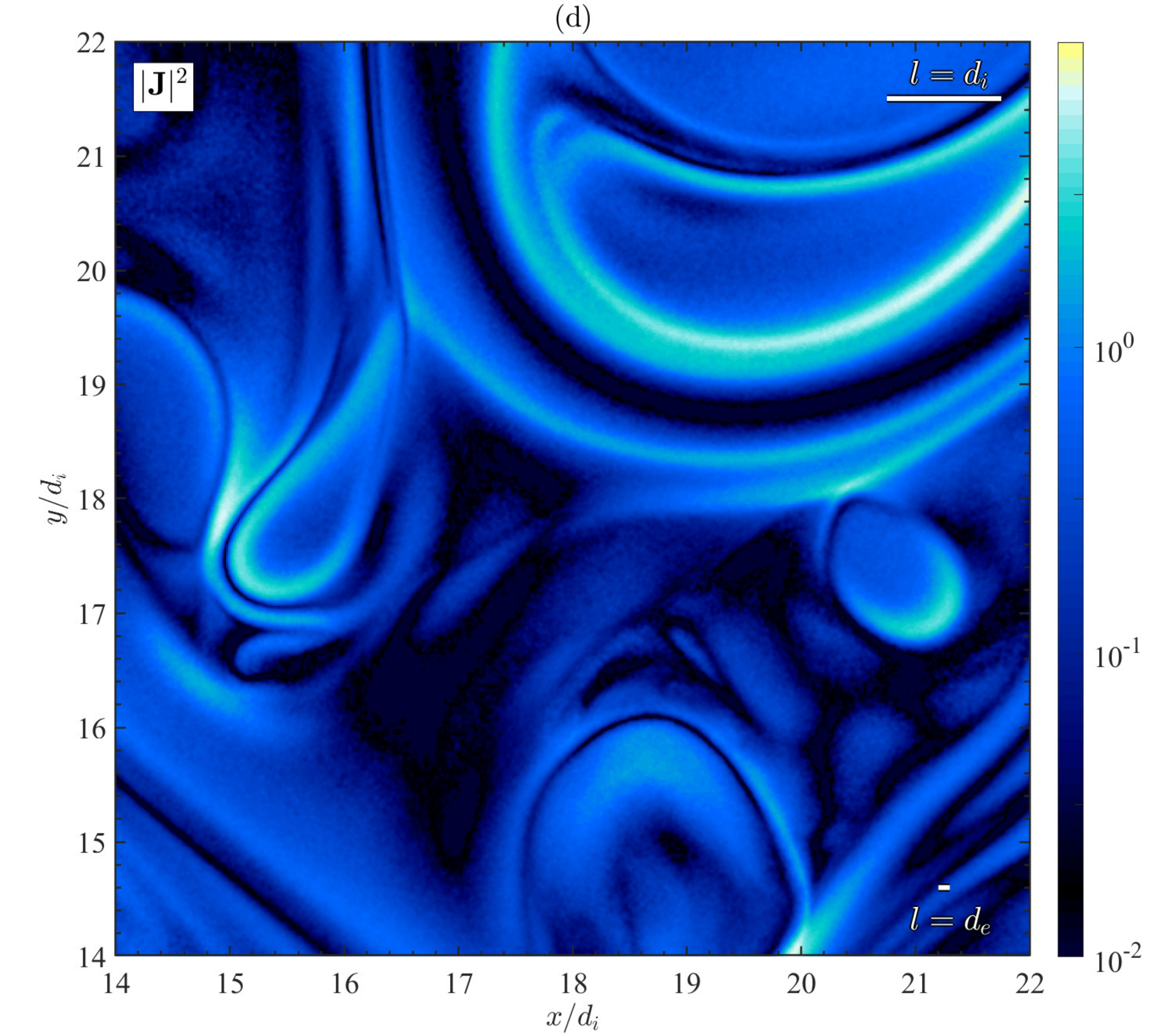}
\caption{Contour plots of the logarithm of the magnitude of the magnetic fluctuations, $|\delta \vect{B}|^2 = |\vect{B}-\vect{B}_0|^2$ \textit{(left column)}, and of the current density, $|\vect{J}|^2$ \textit{(right column)} at $t = t^{peak}$, both in the whole simulation box \textit{(top row)} and in a sub-domain \textit{(bottom row)}. Magnetic vortexes are compared with circles of radius $r = d_{i,e}$ and current sheet widths are compared with lines of length $l = d_{i,e}$.}
\label{fig:dB2}
\end{figure*}

Figure\ref{fig:time_evolutions}(b) shows the time evolution of the maximum of the magnitude of the current density computed over the whole simulation box. This starts increasing after about one $t_{\textrm{NL}}$ and then reaches a first maximum at $t_{\mathrm{rec}} \simeq 4.9\, \Omega_i^{-1}$, which is after about $2\, t_{\textrm{NL}}$. In \citet{Franci_al_2017}, we have provided numerical evidence of the link between the time of the first maximum of $|\vect{J}|$ and the onset of magnetic reconnection events. Here we observe the same, as we see clear signs of reconnection in the magnetic and current structures starting at $t \simeq t_{\mathrm{rec}}$ (not shown). 
Figure~\ref{fig:time_evolutions}(c) shows the rms of the in-plane and out-of-plane components of the magnetic field (red solid and dashed, respectively) and of the ion bulk velocity (light blue). The behaviour of $\vect{B}_\perp^{\textrm{rms}}$ and $\vect{u}_{i,\perp}^{\textrm{rms}}$ is qualitatively the same as previously observed in hybrid simulations (cf. Fig. 1 of \citet{Franci_al_2015b}): after a quick re-adjustment of the initial conditions, the former slightly increases before slowly decreasing, while the latter decreases during the whole evolution. As a result, the initial small excess of magnetic energy further increases reaching a maximum at about half the simulation and then remains almost constant. The parallel components $\delta \vect{B}_\parallel^{\textrm{rms}}$ and $\vect{u}_{i,\parallel}^{\textrm{rms}}$, which are initially zero, quickly start to increase until they reach an almost constant and comparable value, which is a few times smaller than their perpendicular counterparts. This indicates that the levels of compressibility and magnetic compressibility that spontaneously form are relatively small but not negligible. It is interesting to note that $\delta \vect{B}_\perp^{\textrm{rms}}$ reaches a maximum at $t = 5\, \Omega_i^{-1} \simeq t_{\textrm{rec}}$ and starts decreasing when reconnection events start occurring.

Figure~\ref{fig:dB2} shows the isocontours of the energy in the magnetic fluctuations, $|\delta \vect{B}|^2 = |\vect{B} - \vect{B}_0|^2 = \delta \vect{B}_x^2 + \delta \vect{B}_y^2 + \delta \vect{B}_z^2$ (panels (a) and (b)) and in the current density, $|\vect{J}|^2$, both in the whole $32 \, d_i \times 32 \, d_i$ simulation domain (panels (c) and (d)) and in a $8 \, d_i \times 8 \, d_i$ sub-domain (bottom panels). Coherent magnetic structures, i.e., vortexes or islands, are embedded in a more chaotic environment where stretched and twisted shapes emerge. The vortexes have radii from a few times $d_i$ down to a few times $d_e$. Gradients of the magnetic field occur at smaller scales, as the strongest current structures have a width of the order of $d_e$. For an easy comparison by eye, we have also drawn circles with radius $r = d_i$ and $r = d_e$ in fig.~\ref{fig:dB2}(a)-(c) and lines with length $l = d_i$ and $l = d_e$ in fig.~\ref{fig:dB2}(b)-(d). 

\subsection{Spectral properties of the turbulent fluctuations}
\label{subsec:spectra}

\begin{figure*}
\includegraphics[width=0.495\textwidth]{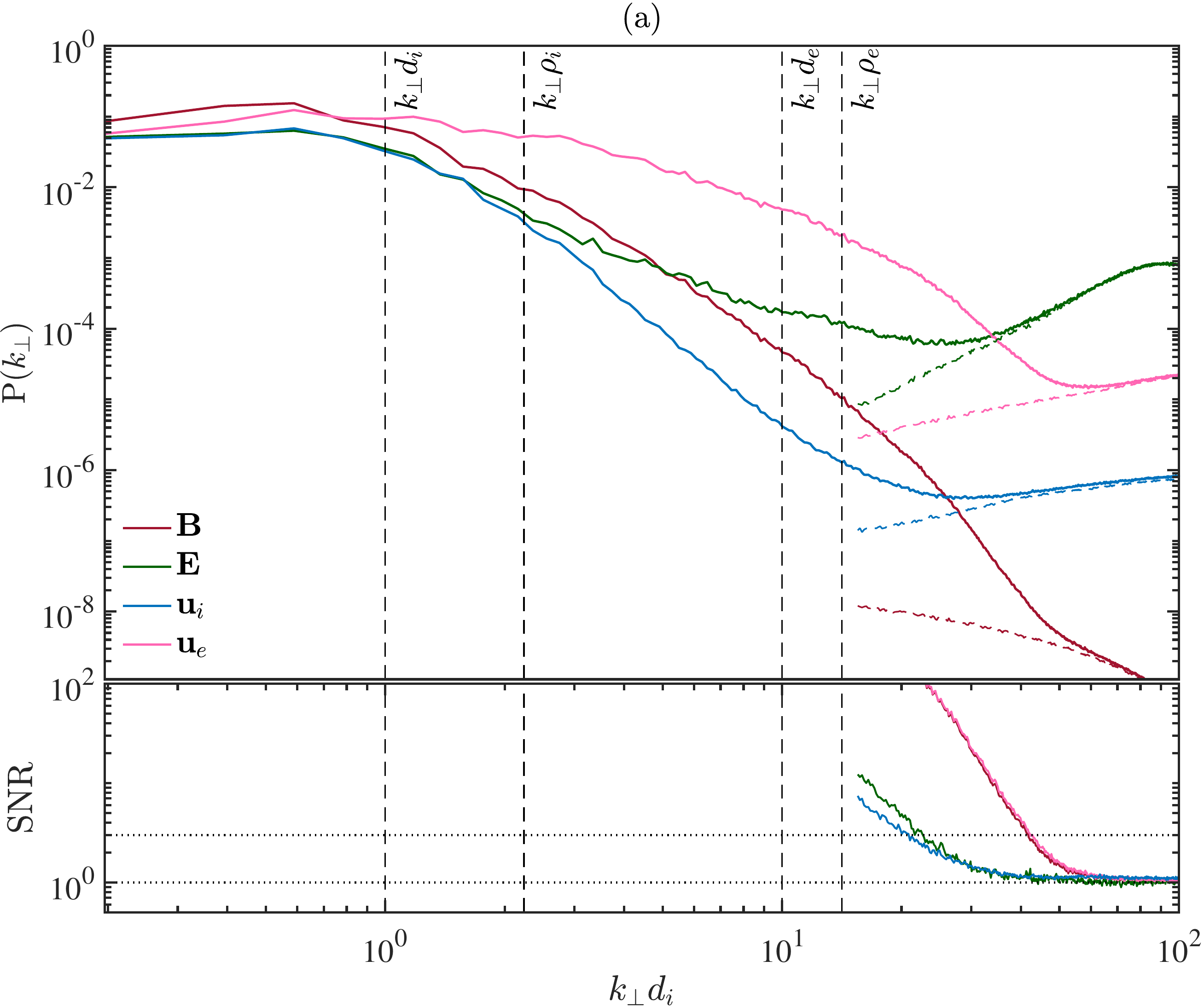}
\includegraphics[width=0.495\textwidth]{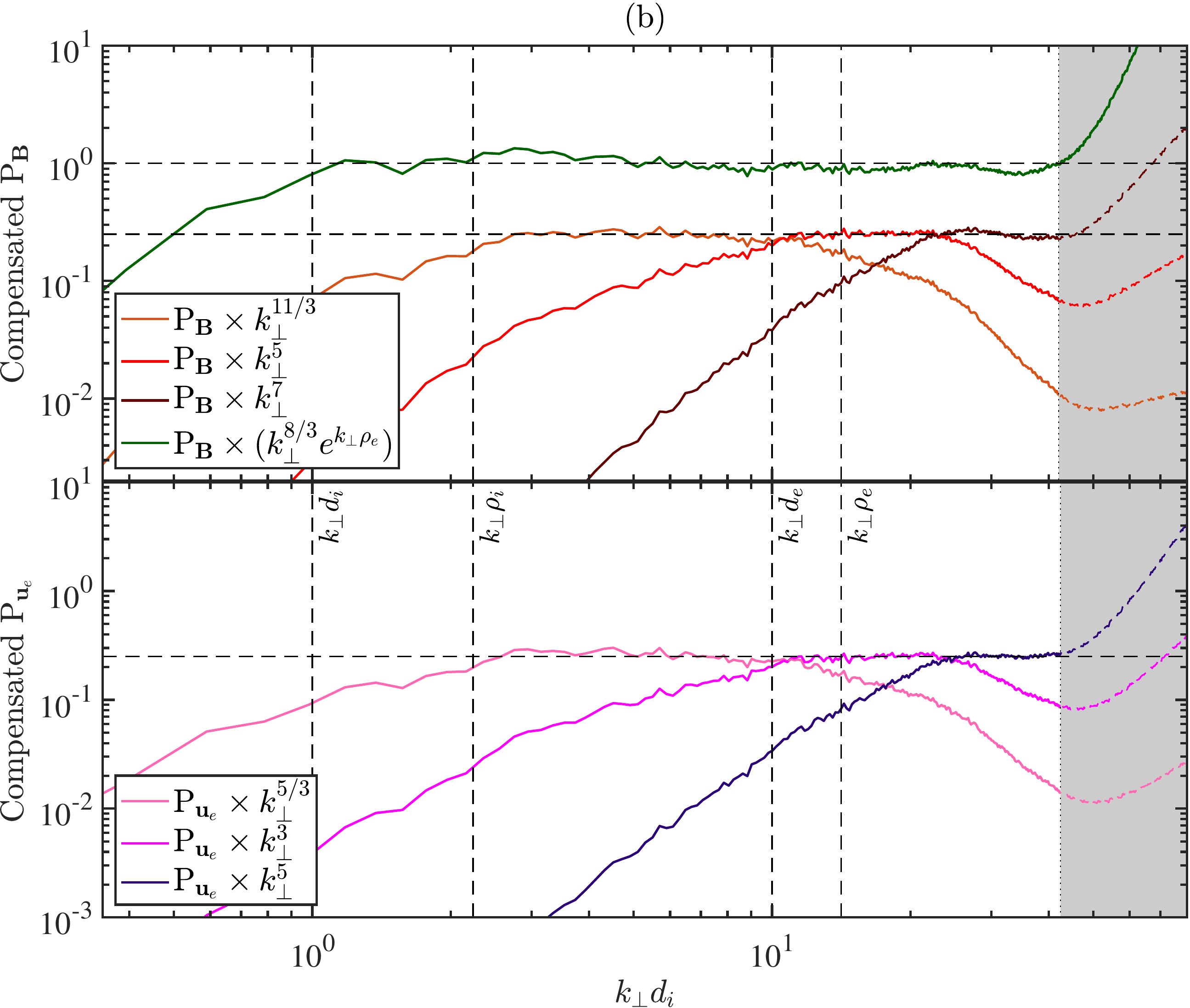}
\caption{Spectral behavior of the  electromagnetic and plasma fluctuations. \textit{Panel (a)}: power spectra of the magnetic field $\vect{B}$, the electric field $\vect{E}$, the ion bulk velocity $\vect{u}_i$, and the electron bulk velocity $\vect{u}_e$ with the respective noise in dashed lines (\textit{top}), and corresponding signal-to-noise ratio SNR (\textit{bottom}). Vertical dashed black lines mark the particle characteristic scales, i.e., the ion and electron inertial length $d_{i,e}$ and gyroradius $\rho_{i,e}$.  \textit{Panel (b)}: power spectra of the magnetic field $\vect{B}$ (\textit{top}) and of the electron bulk velocity fluctuations $\vect{u}_{e}$ (\textit{bottom}), compensated by different power laws (and by an exponential cut-off for $\vect{B}$). The grey area marks the scales at which $\mathrm{SNR} < 3$.}
\label{fig:spectra}
\end{figure*}

Figure~\ref{fig:dB2} has shown that the turbulent structures are randomly oriented, as expected given that the ambient magnetic field is orthogonal to the simulation plane so there is no privileged direction in the plane. We can then reasonably assume the two-dimensional spectra of the turbulent fluctuations to be statistically isotropic and analyse the spectral properties of each field by computing its omnidirectional power spectrum $\mathrm{P}$. 
The top panel of Fig.~\ref{fig:spectra}(a) shows the power spectra of the fluctuations of the magnetic field $\vect{B}$, of the electric field $\vect{E}$, and of the ion and electron bulk velocities $\vect{u}_i$ and $\vect{u}_e$ at $t = t^{peak}$. Black vertical dashed lines mark the wavenumbers corresponding to the ion and electron inertial length, $d_{i,e}$, and gyroradius, $\rho_{i,e}$. For each field, a dashed line with the same color indicate the noise level, estimated by computing the power spectrum at $t = 0$ when only the large-scale initial fluctuations should be present. 
The bottom panel of Fig.~\ref{fig:spectra}(a) shows the signal-to-noise ratio (SNR) for each field. In the following, we will consider the power spectra reliable (meaning that their shape is not affected by the noise) at those scales where $\mathrm{SNR} > 3$. This holds for $k_\bot d_i \gtrsim 20$ (or, equivalently, $k_\bot d_e \gtrsim 2$) for $\mathrm{P}_{\vect{E}}$ and $\mathrm{P}_{\vect{u}_i}$ and for $k_\bot d_i \gtrsim 45$ ($k_\bot d_e \gtrsim 4.5$) for for $\mathrm{P}_{\vect{B}}$ and $\mathrm{P}_{\vect{u}_e}$. This ensures that our modelling of the magnetic field power spectrum is meaningful around the electron scales and for a further half a decade in wavenumber above $k_\bot d_e = 1$.
The top panel of Figure~\ref{fig:spectra}(a) suggests that both $\mathrm{P}_{\vect{B}}$ and $\mathrm{P}_{\vect{u}_e}$ may exhibit a power-law behaviour with different spectral indices in different ranges above and below the electron scales. On the contrary, $\mathrm{P}_{\vect{u}_i}$ starts dropping at $k_\bot d_i \gtrsim 1$ and its behavior is not so clear. There is a hint that for $4 \lesssim k_\bot d_i \lesssim 10$ it may exhibit a similar slope to the magnetic field (although such interval is too limited to make any claim) and then it starts flattening just before the noise level is reached, so it is not possible to infer whether this is a numerical or physical effect. Finally, $\mathrm{P}_{\vect{E}}$ at sub-ion scales is much less steep than $\mathrm{P}_{\vect{B}}$, so that the level of the electric field fluctuations is much larger than the magnetic field's for $k_\bot d_i \gtrsim 5$. 
In order to provide a more quantitative characterization of $\mathrm{P}_{\vect{B}}$ and $\mathrm{P}_{\vect{u}_e}$, Fig.~\ref{fig:spectra}(b) shows them compensated by different powers of $k_\bot$. The top panel of Fig.~\ref{fig:spectra}(b) shows that the magnetic field power spectrum can be modelled by $\mathrm{P}_{\vect{B}} \propto k_\bot^{\alpha_{B}}$ with $\alpha_{B}\simeq -11/3$ for almost a decade at $k_\bot d_e <1$ and $-5$ for a shorter interval at $k_\bot d_e >1$. Correspondingly, we observe $\mathrm{P}_{\vect{u}_e} \propto k_\bot^{\alpha_{e}}$ with $\alpha_{e} \simeq -5/3$ and $-3$ in the two intervals of scales, respectively. The relation $\alpha_{e} = \alpha_{B} + 2$ can be easily explained by considering the definition of the current density, $\vect{J} = n_i \vect{u}_i - n_e \vect{u}_e$, and Amp\`ere's law which links it to magnetic field, $\vect{J} = \nabla \times \vect{B}$ (where the displacement current has been neglected). Due to charge neutrality, $n_i = n_e = n$ and, assuming that $n = n_0 + \delta n$ with $\delta n \ll n_0$, we get $\vect{u}_e \propto \nabla \times \vect{B}$. In Fourier space this reads $\vect{u}_{e,k} \propto \vect{k} \times \vect{B}_k$, from which we obtain $\mathrm{P}(\vect{u}_e) \propto k_\bot^2 \mathrm{P}(\vect{B})$ (since in 2D $\vect{k} = \vect{k}_\bot$).
At smaller scales, $k_\bot d_e \gtrsim 2.5$, there is a hint of another possible power-law range in $\mathrm{P}(\vect{u}_e)$, with$\alpha_{e} \simeq -5$. This interval is very narrow, just about a factor of 2 in wavenumber, as the power spectrum quickly reaches the noise level. Although we cannot consider this power-law behavior and its slope reliable enough, the further steepening of the power spectrum is evident from the top panel of Fig.~\ref{fig:spectra}(a). In the same interval, the magnetic field power spectrum also steepens accordingly, although $\alpha_{e} = \alpha_{B} + 2$ does not seem to hold perfectly anymore, possibly due to the effect of the large noise in the density due to which the assumption $\delta n \ll n_0$ might not hold anymore and/or to the fact that at large wavenumbers (and frequencies) we cannot neglect the displacement current anymore. The fact that at the smallest scales $\mathrm{P}(\vect{u}_e)$ seems to exhibit still a power-law behavior $\mathrm{P}(\vect{u}_e)$ while $\mathrm{P}(\vect{B})$ seems to follow approximately, might represent a hint that the turbulent cascade is still proceeding at $k_\bot \rho_e \gtrsim 1$, but the dynamics might be dominated by the electron processes rather than by the magnetic fluctuations. Investigating this will require new simulations with an even higher resolution and larger number of electrons per cell, in order to further improve the accuracy of the power spectra below the electron scales.

As mentioned in Sec.~\ref{sec:introduction}, there is no consensus in the literature on the shape of the magnetic field power spectrum across the electron characteristic scales and, more precisely, on whether it exhibits a double power-law behavior or rather an exponential decay. So far, we have discussed how two power laws with a spectral index of $-11/3$ and $-5$ seems to provide a good modelling for $P_\vect{B}$ over a little more than a full decade across the electron scales, more specifically for $0.25 \lesssim k_\bot d_e \lesssim 4$. We also compensated the magnetic field power spectrum by the inverse of an exponential cut-off similar to that suggested by ~\citet{Alexandrova_al_2021}, which reads $P_\vect{B} \propto k_\bot^{-8/3} \exp{(- c \,k_\bot \rho_e)}$ with $c = 1.8$. The result is shown by the green line in the top panel of Fig.~\ref{fig:spectra}(b). This also seems to provide a good approximation for $P_\vect{B}$ as it is almost a horizontal line, although with significant oscillations, for a little more than a decade in wavenumber. We need to note, however, that we have set $c = 1$ instead of $c = 1.8$ as in ~\citet{Alexandrova_al_2021}, since the latter does not work well here. The difference in this constant might be due either to the unrealistic mass ratio employed in the simulation or to the fact that the exponential cut-off model fails in our case, possibly due to the specific plasma conditions. Providing strong numerical evidence in favour or against this goes beyond the scope of this paper, as we only intended to show that the simulation is able to model the turbulent cascade accurately up to scales smaller than the electron gyroradius, regardless of what the exact shape of the power spectra is.

\begin{figure*}
\includegraphics[width=0.48\textwidth]{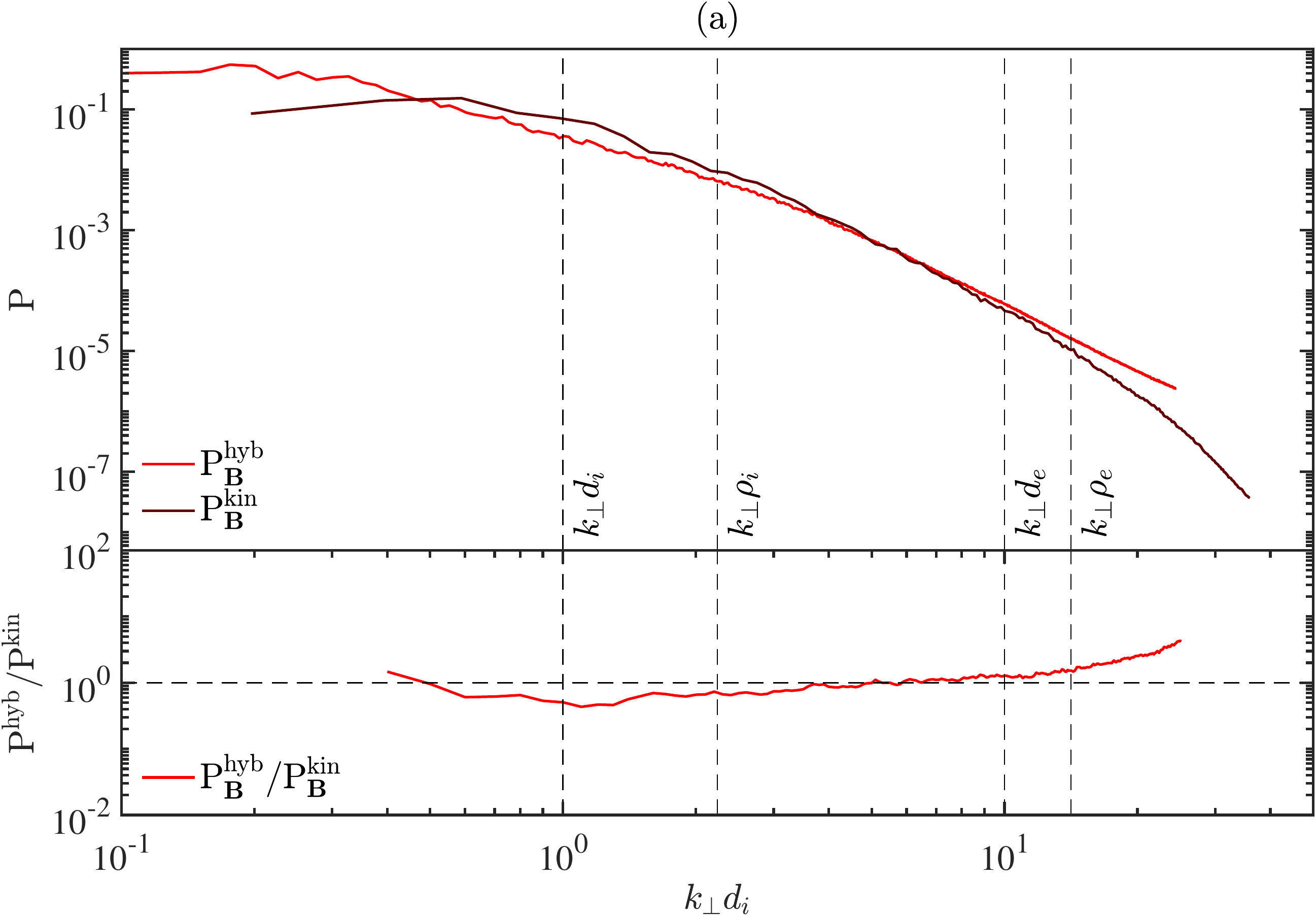}
\includegraphics[width=0.48\textwidth]{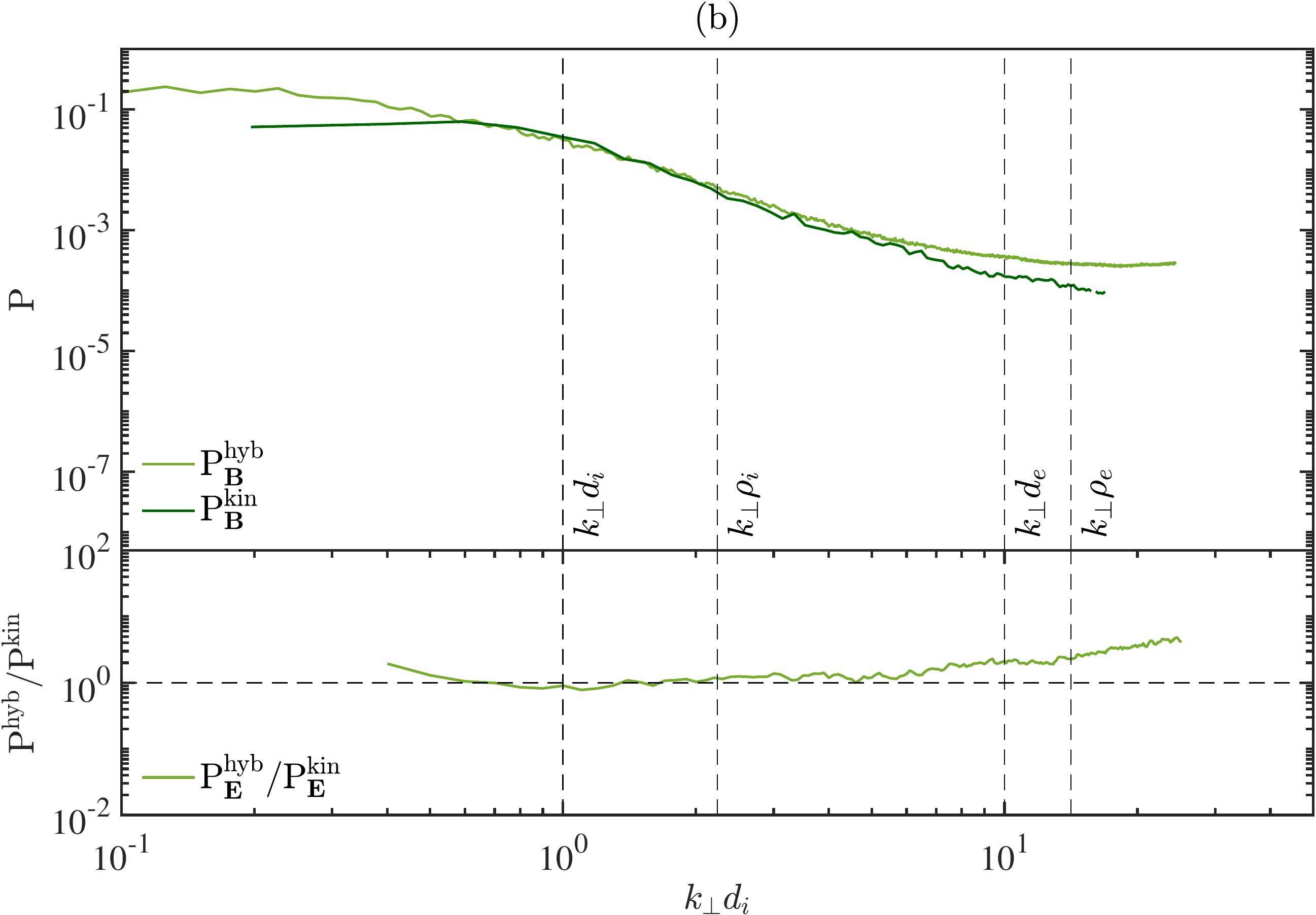}\\
\includegraphics[width=0.48\textwidth]{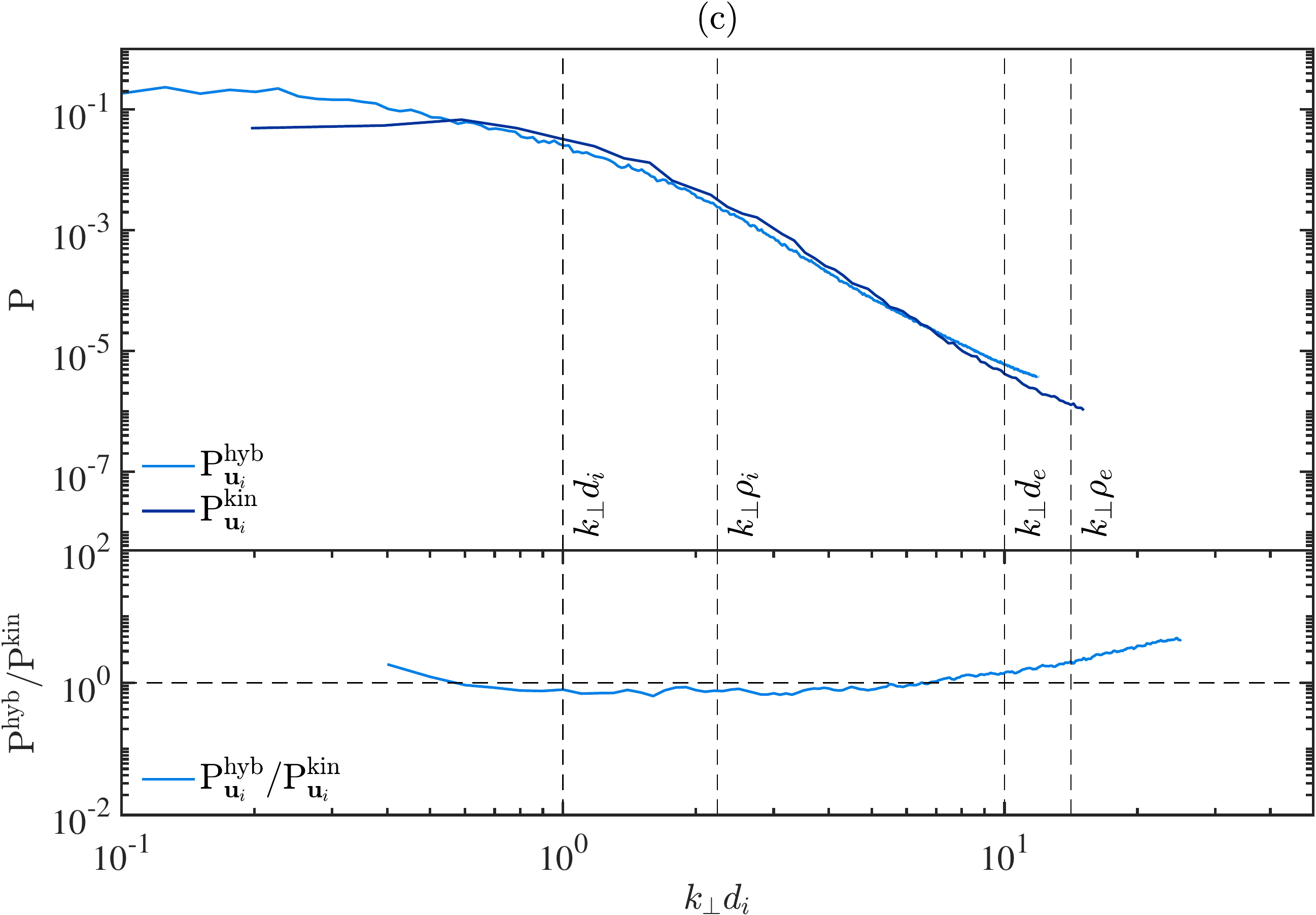}
\includegraphics[width=0.48\textwidth]{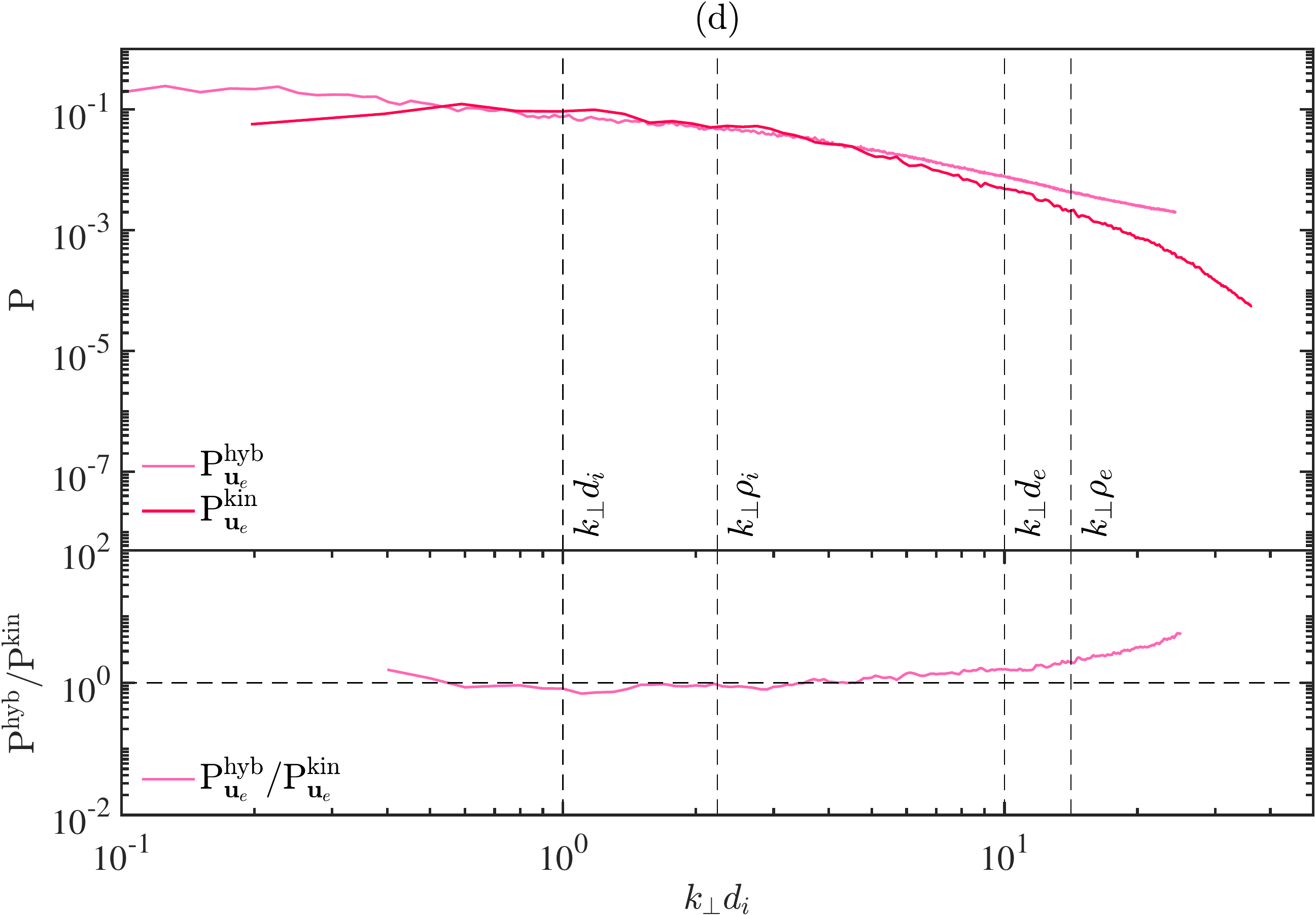}\\
\includegraphics[width=0.48\textwidth]{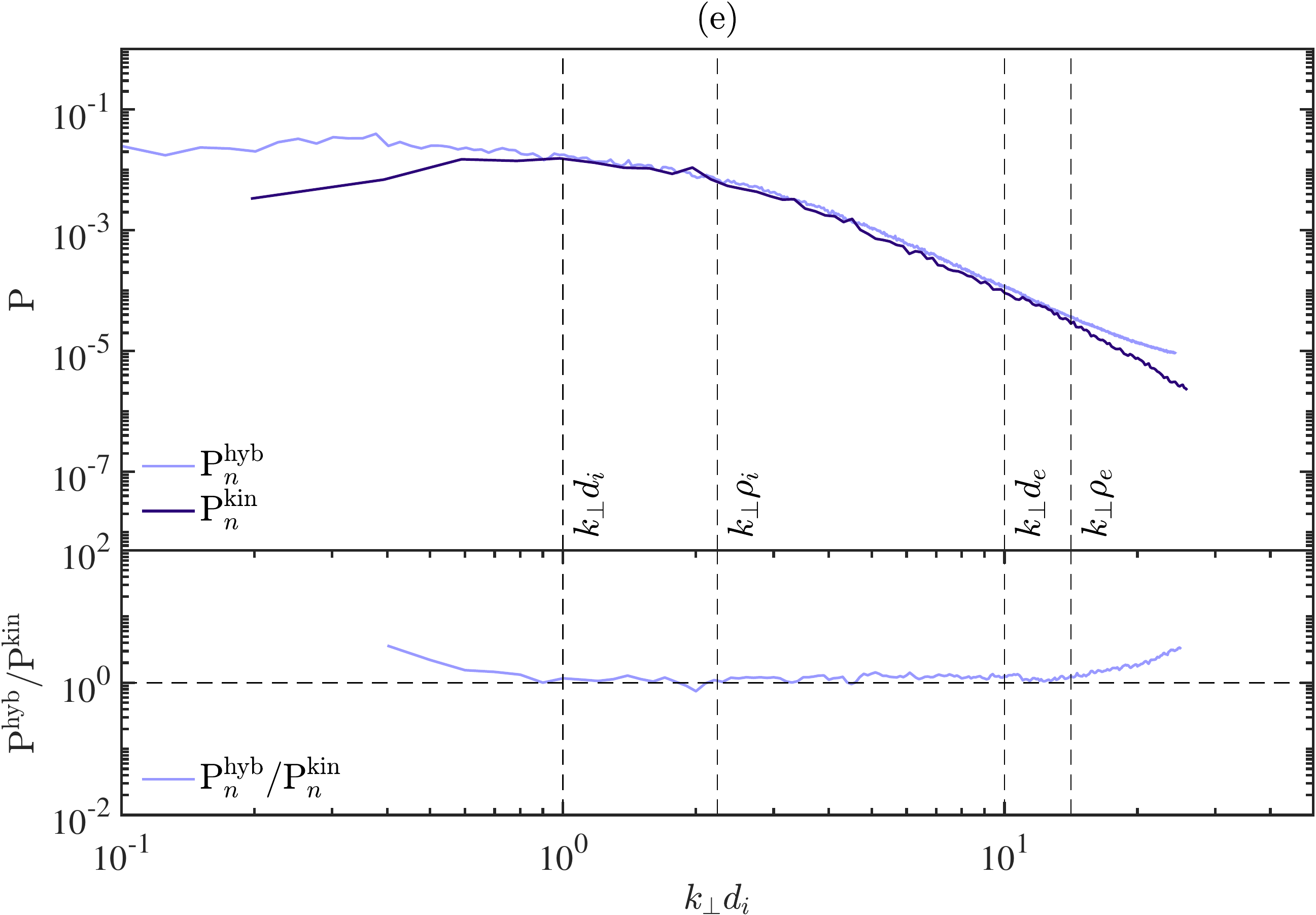}
\includegraphics[width=0.48\textwidth]{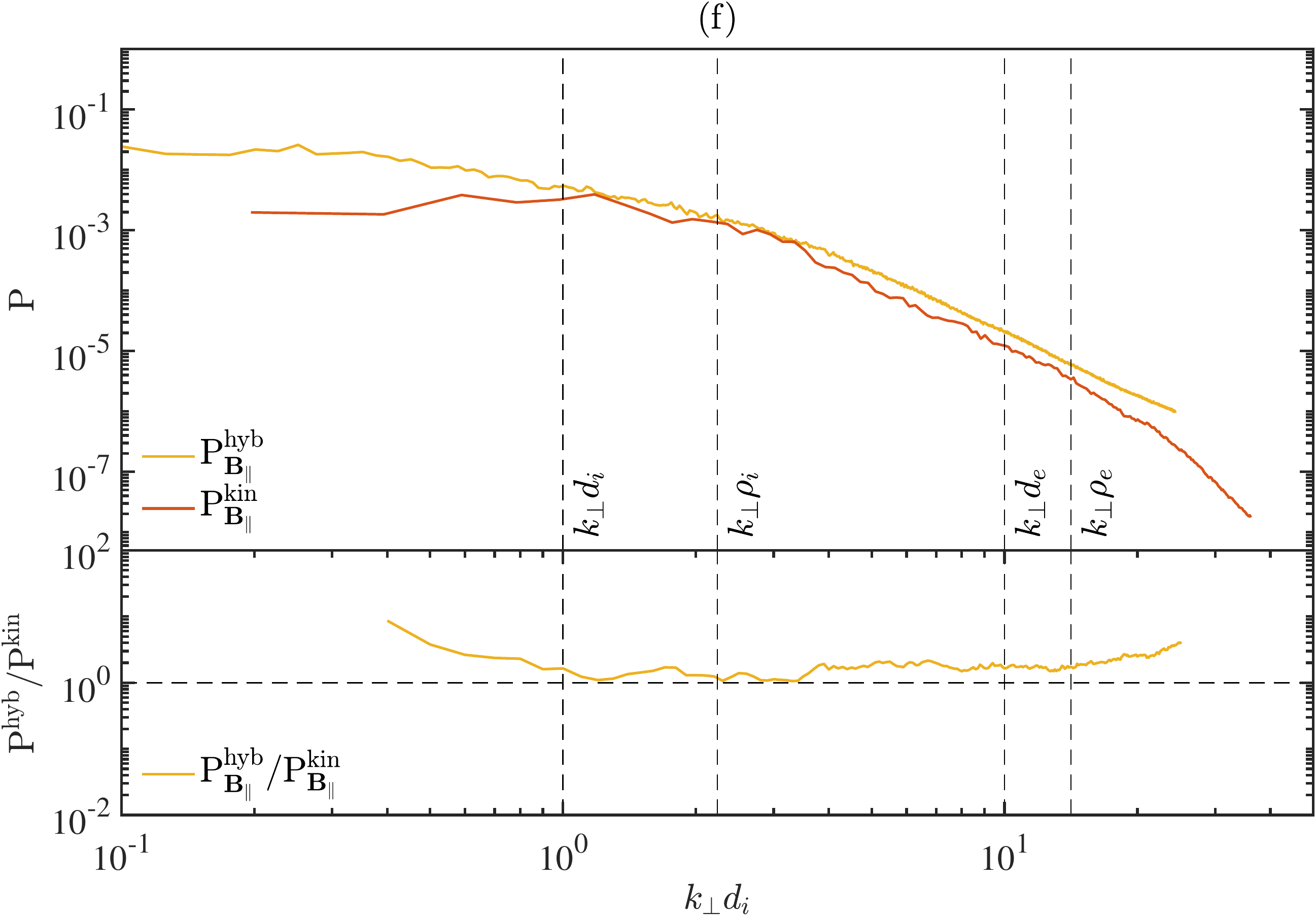}
\caption{Direct comparison between the power spectra of the kinetic simulation, $\mathrm{P}^{\mathrm{kin}}$, and those of the hybrid simulation $\mathrm{P}^{\mathrm{hyb}}$ for the magnetic field $\vect{B}$ (a), the electric field $\vect{E}$ (b), the ion bulk velocity $\vect{u}_i$ (c), the electron bulk velocity $\vect{u}_e$ (d), the ion density $n_i$(e), and the parallel magnetic field $\vect{B}_\parallel$ (f). For each field, the ratio $\mathrm{P}^{\mathrm{hyb}}$/$\mathrm{P}^{\mathrm{kin}}$ is also shown in the bottom panels.}
\label{fig:picVShybrid}
\end{figure*}

A limitation of this fully kinetic simulation, as for most simulations employing the same model, is the fact that the MHD scales are not retained. Usually, especially when explicit numerical codes are used, this is due to the fact that all characteristic spatial scales need to be resolved, including the Debye length. This is not a requirement in our case, as the iPic3D code employs the implicit moment method. Here, however, the box size is still limited as we decided to prioritize employing a very large number of ppc to reduce the numerical noise. While this choice assures that the power spectra are more accurate at large wavenumbers, missing completely the MHD range makes the turbulent cascade start developing at the ion scales. Some important properties (e.g., the plasma and magnetic compressibility, the residual energy) at ion scales are set by our initial conditions rather than self-developing from the larger scales. One could then wonder whether this sub-ion scale turbulence is representative of the plasma dynamics in a larger plasma system or if it is instead sensibly constrained or affected by the box size. In order to investigate this and validate our results, we have run a hybrid simulation with the same physical conditions but a much larger box, able to model the turbulent cascade over two decades in wavenumber, one above and one below the ion scales. This simulation has the same parameters and initial conditions as the one described in~\citep{Franci_al_2020arxiv}, so that further information can be found in Appendix A therein. The only (minor) differences are that the grid points are $4000 \times 4000$ instead of $4096 \times 4096$ (and correspondingly the box size is $250 \, d_i$ instead of $256$) and $2048$ ppc instead of $1024$. 

In Fig.~\ref{fig:picVShybrid} we compare the power spectra of different fields for the fully kinetic simulation, $\mathrm{P}^{\mathrm{kin}}$, and for the hybrid simulation, $\mathrm{P}^{\mathrm{hyb}}$. For the former, we only draw the portion of the power spectra where the corresponding SNR is above 3. For the latter, since we have only the initial power spectra for $\vect{u}_i$ and $n_i$, we apply the SNR criterium by using the noise threshold $\mathrm{SNR}({\vect{u}_i})<3$ for $\mathrm{P}_{\vect{u}_i}$ and $\mathrm{SNR}(n_i)<3$ for all other power spectra.
Fig.~\ref{fig:picVShybrid}(a) compares the spectral behavior of the magnetic fluctuations. $P_{\vect{B}}^{\mathrm{hyb}}$ is more extended at large scales, since the hybrid simulation box is larger. This allows for the development of a Kolmogorov-like power law with a spectral index of $-5/3$ for $k_\bot \d_i \lesssim 3.5$, which then steepens to a power law with a slope compatible with $-11/3$, as already observed in~\citet{Franci_al_2020arxiv}. $P_{\vect{B}}^{\mathrm{kin}}$, instead, completely lacks the $-5/3$ range and is higher at the largest scales. This is likely due to two main reasons. First, the injection scale is just in the middle of that range, and probably the fluctuations at the largest scales did not have time to fully partake in the cascade. Secondly, in the fully kinetic simulation the mechanisms leading to a fully developed inertial range are missing and therefore those underlying the sub-ion scale turbulence may start determining the dynamics at slightly larger scales. Between the ion-scale break of the hybrid simulation (the end of MHD inertial range) and the electron scales, i.e., in the range $3.5 \lesssim k_\bot d_i \lesssim 10$, $P_{\vect{B}}^{\mathrm{hyb}}$ and $P_{\vect{B}}^{\mathrm{kin}}$ overlap. Finally, at $k_\bot d_i \simeq 10$, i.e., $k_\bot d_e \sim k_\bot \rho_e \sim 1$, when $P_{\vect{B}}^{\mathrm{kin}}$ steepens as the electron physics kicks in, $P_{\vect{B}}^{\mathrm{hyb}}$ maintains the same slope until the noise level is reached. In Fig.~\ref{fig:picVShybrid}(b), $P_{\vect{E}}^{\mathrm{hyb}}$ and $P_{\vect{E}}^{\mathrm{kin}}$ are compared. These are quite overlapped in the range $0.6 \lesssim k_\bot d_i \lesssim 6$ while, at smaller scales, the former flattens while the latter keeps going down with a more or less constant slope. This difference in behavior may be partially due to the different level of noise in the two simulations, as $\Delta x^{\mathrm{hyb}} = 4 \, \Delta x^{\mathrm{kin}}$ and $\mathrm{ppc}^{\mathrm{hyb}} = 1/4 \, \mathrm{ppc}^{\mathrm{kin}}$. The latter condition is related to the fact that in the hybrid case the electrons are treated as an isothermal fluid, so that $P_e \propto \beta_e/2 \, n_i$ and the number of ions per cell (2048) determines the noise in the electric field, while in fully kinetic case $P_e \propto n_e T_e$, so the number of electrons per cell (8192) is the determining one. As a consequence, a sub-domain of a given size contains 16 times more particles in the fully kinetic simulation with respect to the hybrid one. It is, however, reasonable to ascribe the flattening at scales much larger than the spatial resolution to different physics, given that in the hybrid model we approximate $\bnabla \cdot P_e \sim \beta_e/2 \bnabla n_i$, so any possible effect due to electron temperature anisotropy and non-gyrotropy, which can be expected to become important when the electron scales are approached, is not retained in the hybrid model. 
In Fig.~\ref{fig:picVShybrid}(c) we compare the power spectra of the ion bulk velocity. These are very close to each other at all the scales where the noise is negligible, with just a minor difference when reaching the electron scales. From this comparison, we can conclude that the spectral behavior of the ion velocity seems to be the same in the two simulations, thus it is not significantly affected by the presence of kinetic electrons and related processes. The power spectra of the electron bulk velocity, shown in Fig.~\ref{fig:picVShybrid}(d), start overlapping at $k_\perp d_i \simeq 0.5$ and they proceed together for about a decade, up to $k_\perp d_e \simeq 0.5$, where $P_{\vect{u}_e}^{\mathrm{hyb}}$ further steepens, accordingly with $P_{\vect{B}}^{\mathrm{hyb}}$ as explained above.
Figure~\ref{fig:picVShybrid}(e) compares the power spectra of the ion density fluctuations. We note here that for the fully kinetic simulation we have verified that the spectrum of the ion density, $P_{n_i}$, and that of the electron density, $P_{n_e}$, are exactly identical up to $k_\perp d_i \simeq 40$, where they start being affected by the two different levels of noise (due to the different number of pcc for the two species).
We observe that $P_{n_i}^{\mathrm{hyb}}$ and $P_{n_i}^{\mathrm{kin}}$ almost perfectly overlap in the whole range $k_\bot d_i \simeq 1$ and $k_\bot \rho_e \simeq 1$, with $P_{n_i}^{\mathrm{hyb}}$ getting larger just below the electron scales.
Finally, Fig.~\ref{fig:picVShybrid}(f) compares the power spectra of the parallel magnetic fluctuations.
Interestingly, $P_{\vect{B}_z}^{\mathrm{hyb}}$ is slightly larger than $P_{\vect{B}_z}^{\mathrm{kin}}$ at all scales, especially below the ion-scale break where they exhibit the same spectral slope therefore keeping a constant differing factor. This could be somehow related to the different initial amplitude of the turbulent fluctuations with respect to the ambient field, given that the ratio $P_{\vect{B}_z}^{\mathrm{hyb}}/P_{\vect{B}_z}^{\mathrm{hyb}}$ is compatible with the ratio $\vect{B}_{\mathrm{rms}}^{\mathrm{hyb}} /\vect{B}_{\mathrm{rms}}^{\mathrm{kin}} \simeq 1.2$. Another possible reason for the difference in the magnetic compressibility could be related to the plasma betas: although we start the two simulations with the same values of $\beta_i$ and $\beta_e$, these evolve differently and reach different values when the turbulence is fully developed (in particular, $\beta_e$ does not change in the hybrid case by definition).
This difference, however, is very small and as such cannot be consider as an indication of a different regime or different underlying physics. 

Fig.~\ref{fig:picVShybrid} globally returns a picture where the MHD-scale fluctuations and their properties and evolution do not seem to have significant effects on those below the ion scales. This is supported in particular mode by the spectral behaviour of the density fluctuations, since these are zero at the beginning (apart from the small-scale noise due to the finite number of ppc) and only develop self-consistently through plasma processes. This further validates the findings by \citet{Franci_al_2020} on the kinetic turbulent cascade: the plasma dynamics in the kinetic range is mainly controlled by a few fundamental physical parameters, so that the properties of the fluctuations at sub-ion scales are independent of the presence and extension of an inertial range. 

Since in the fully kinetic simulation $\beta_e = 0.5$, then $\rho_e = \sqrt{\beta_e} \, d_i \simeq 0.7 \, d_i $ so the two electron characteristic scales are very close to each other. It is not possible to infer whether one of the two, or a combination of them, determines the steepening of the power spectra around the electron scales. Preliminary analysis on another fully kinetic simulation (not shown here) with $\beta_e = 0.04$, where $\rho_e = 0.2 d_e$, seems to suggest that the electron-scale break is related to $k_\bot \rho_e$ rather than to $k_\bot d_e$. At this level, however, we cannot exclude that the break might occur at either of the two scales depending on the value of $\beta_e$, just like it is observed to happen for the role of $\beta_i$ for the ion-scale break~\citep{Chen_al_2014b,Franci_al_2016b}.

\subsection{Anisotropic ion and electron heating}
\label{subsec:heating}

\begin{figure*}
\centering
\includegraphics[width=0.42\textwidth]{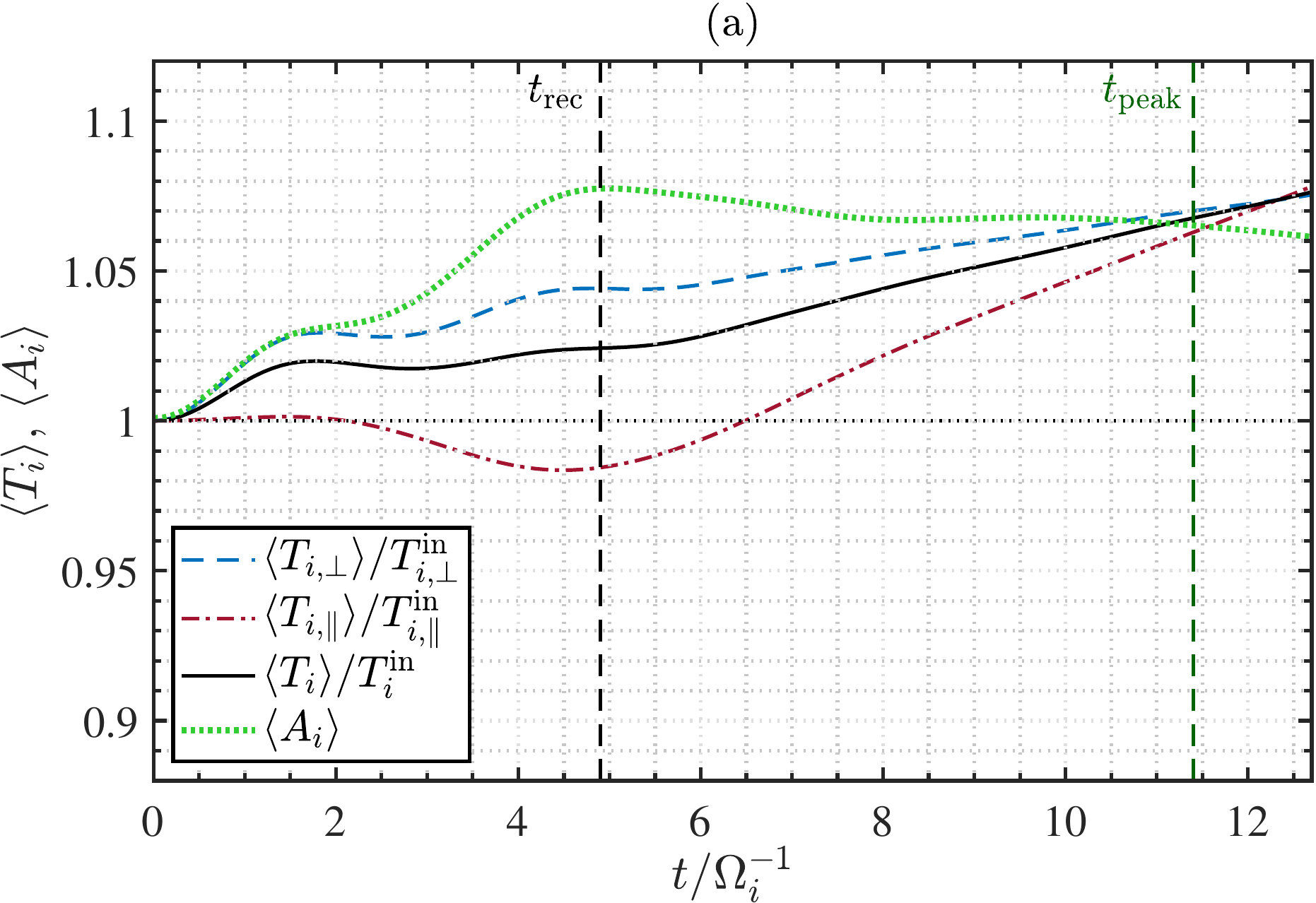}
\includegraphics[width=0.42\textwidth]{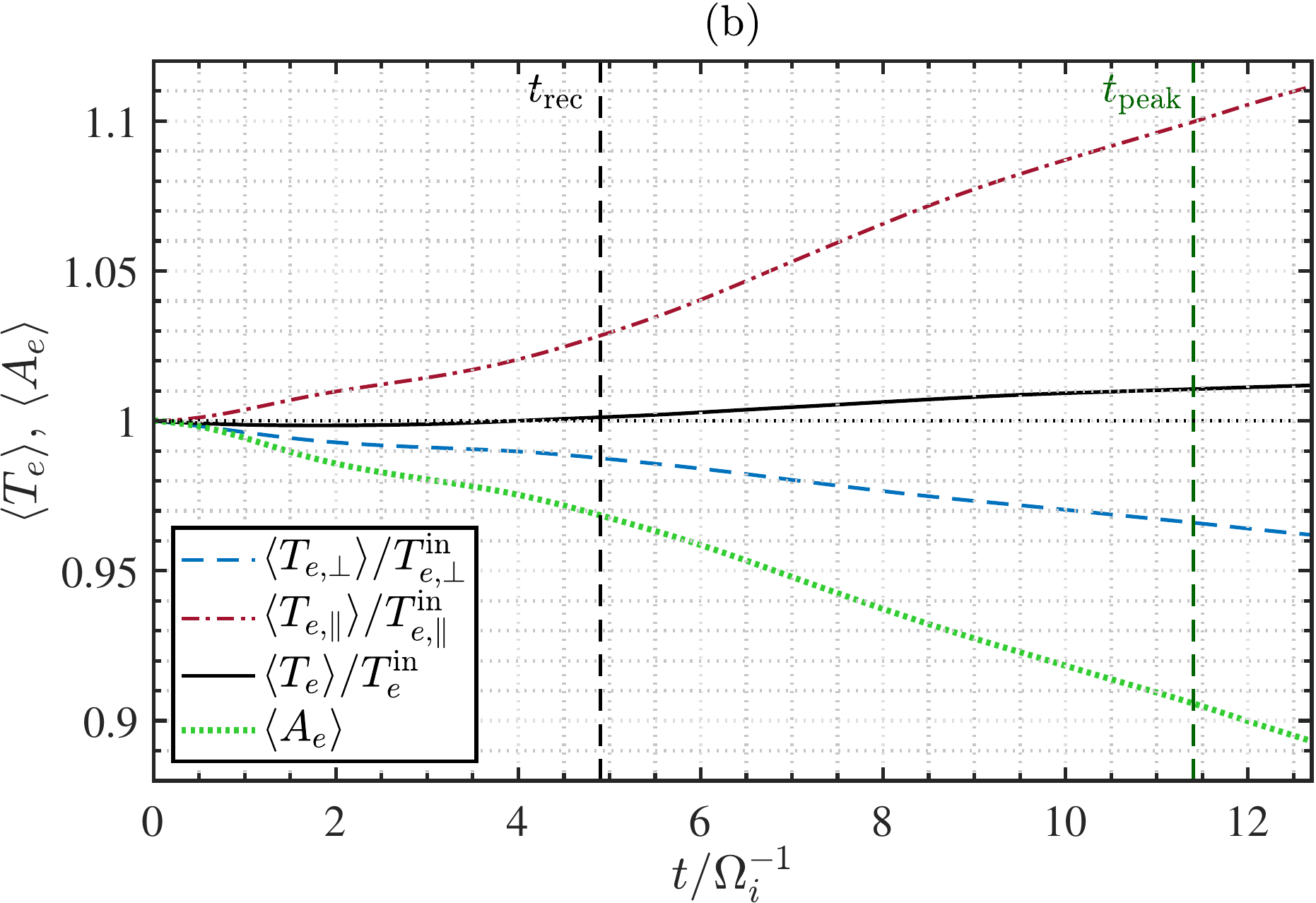}\\
\includegraphics[width=0.325\textwidth]{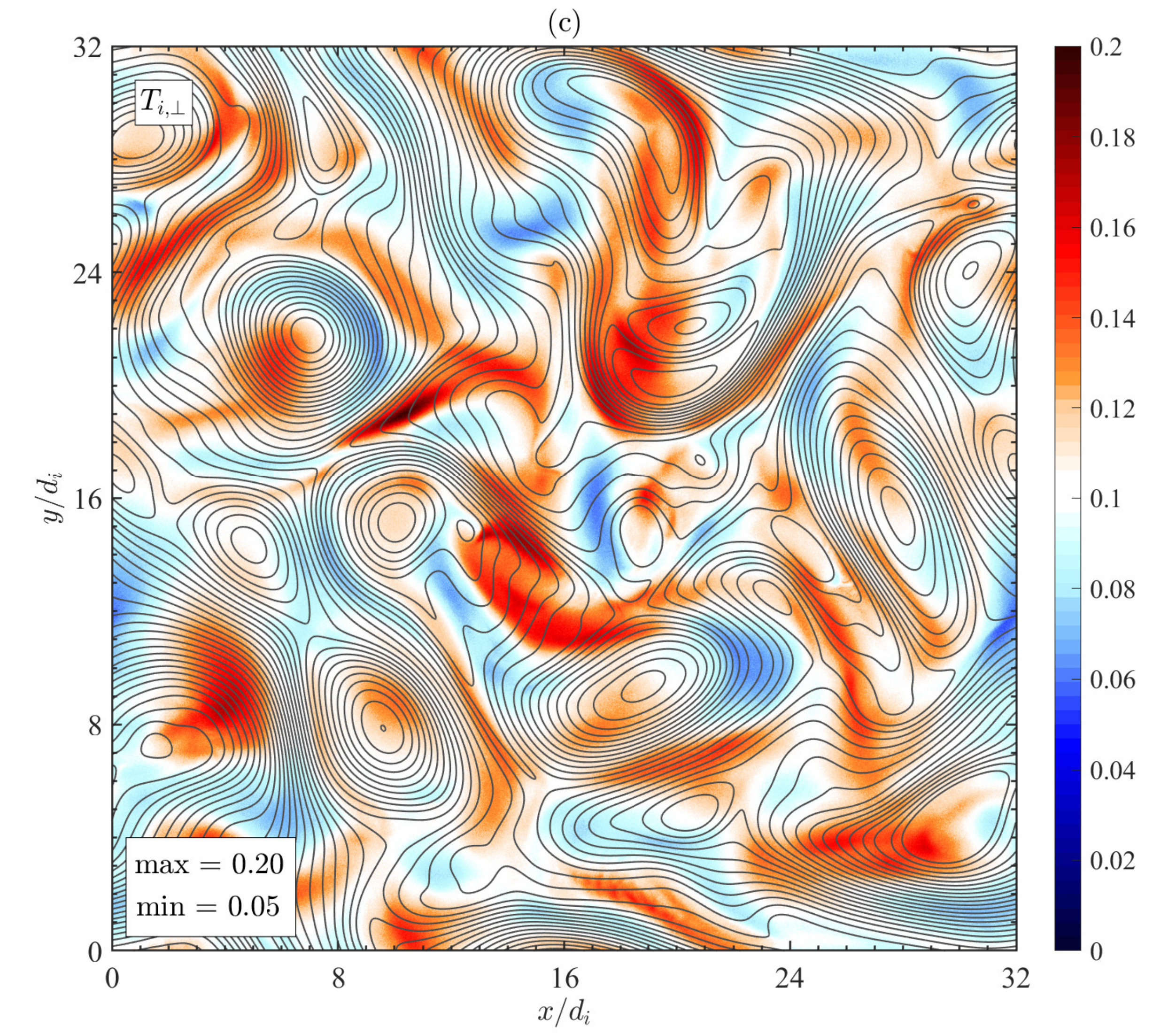}
\includegraphics[width=0.325\textwidth]{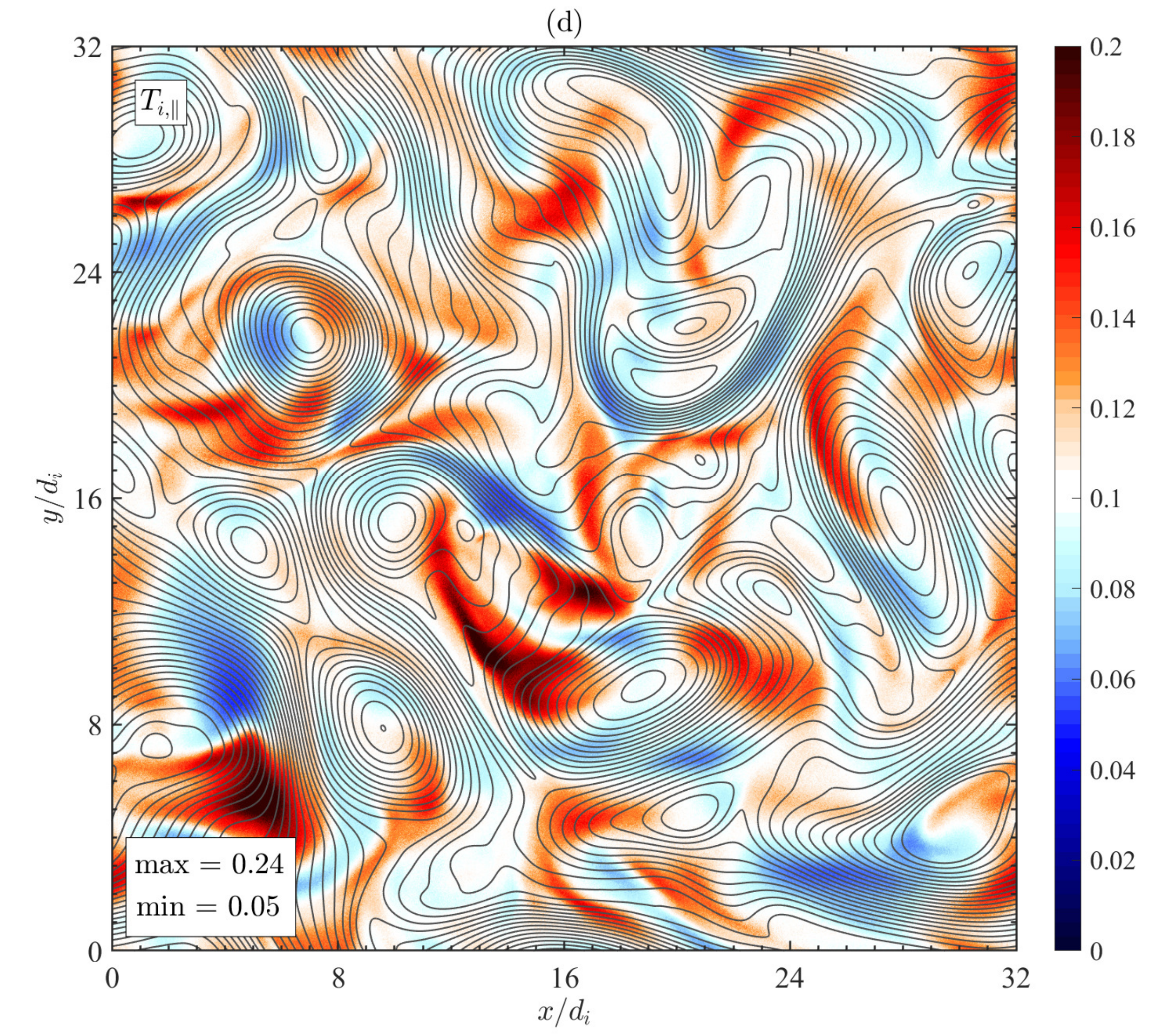}  
\includegraphics[width=0.325\textwidth]{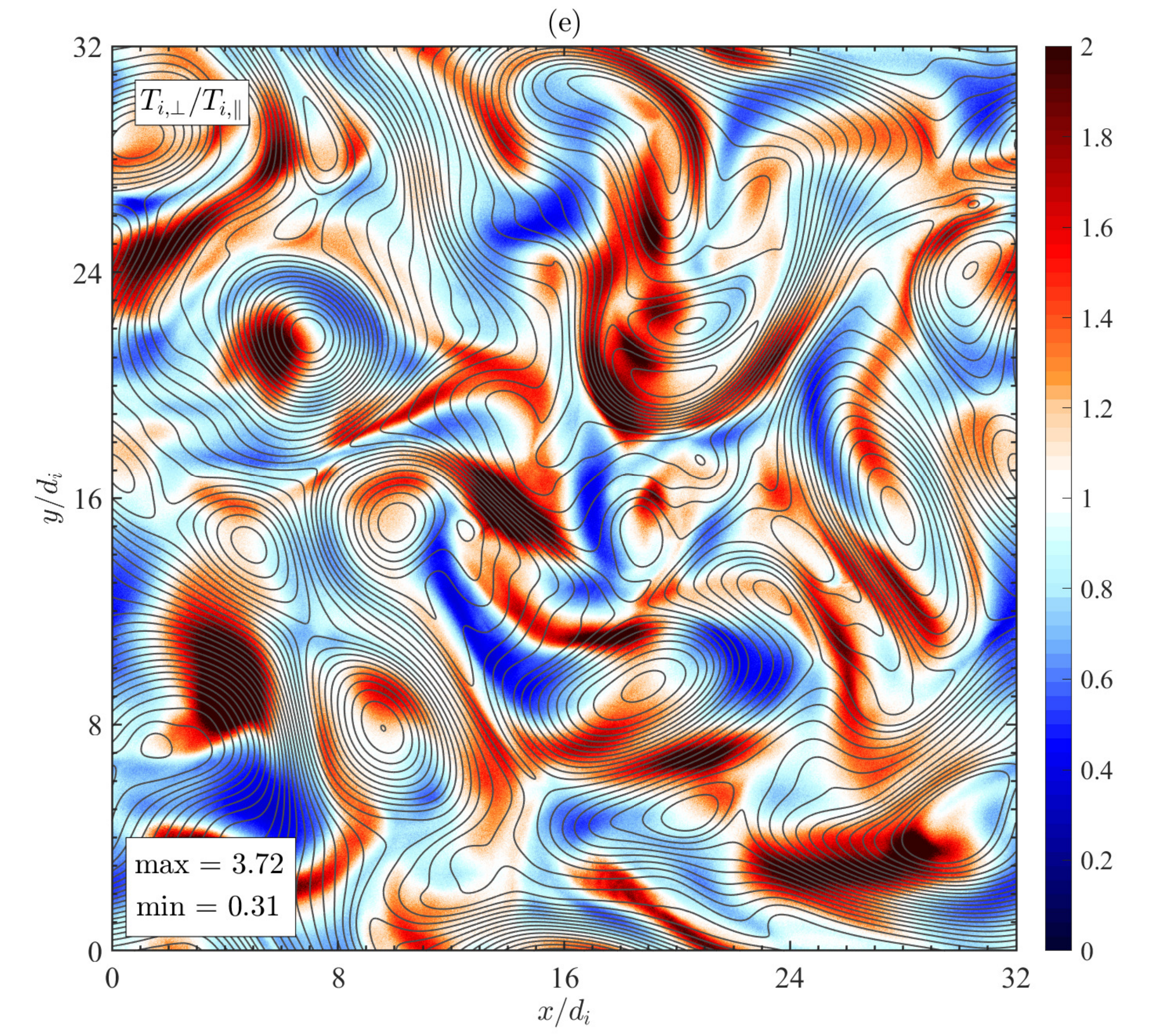}\\
\includegraphics[width=0.325\textwidth]{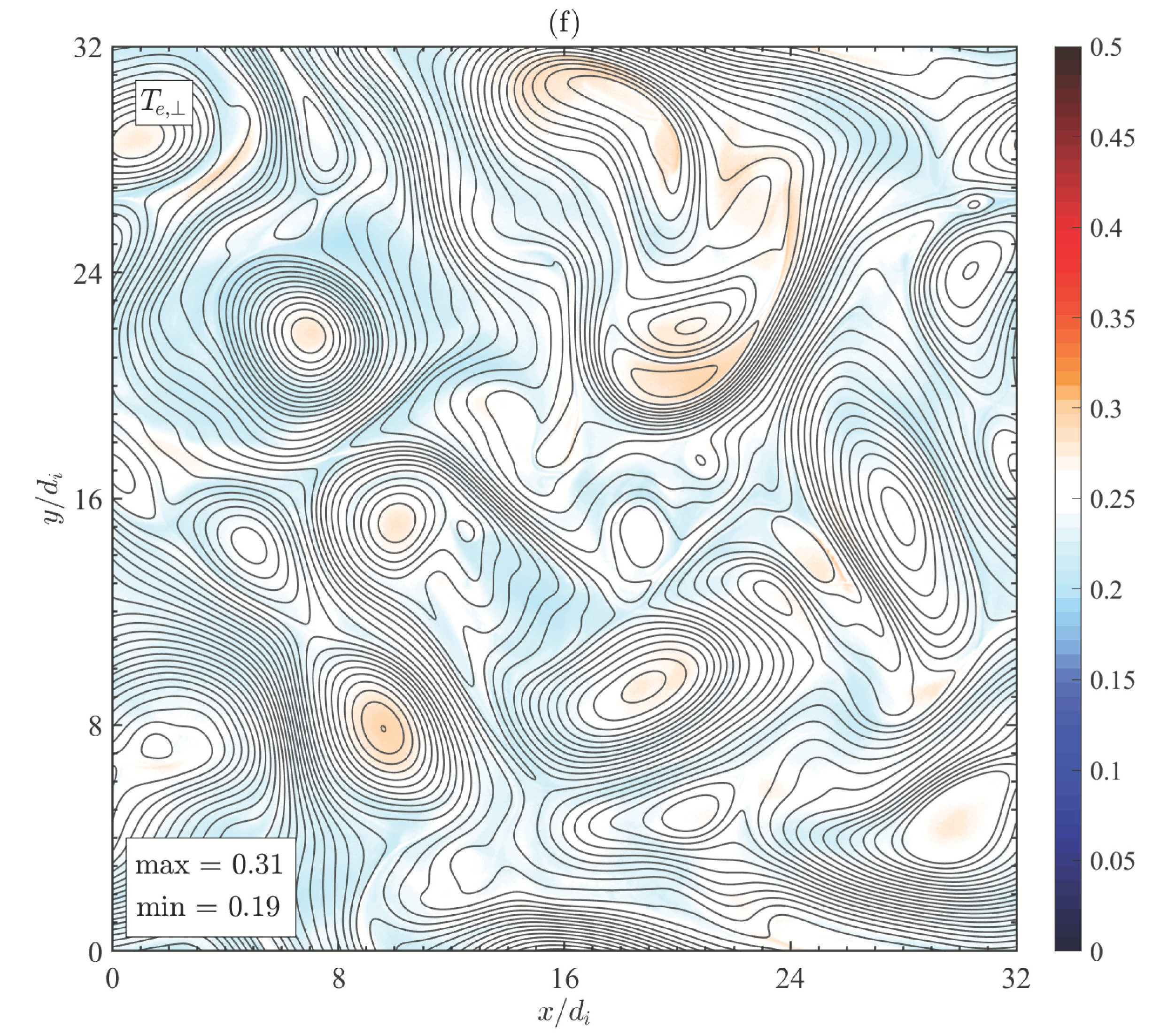}
\includegraphics[width=0.325\textwidth]{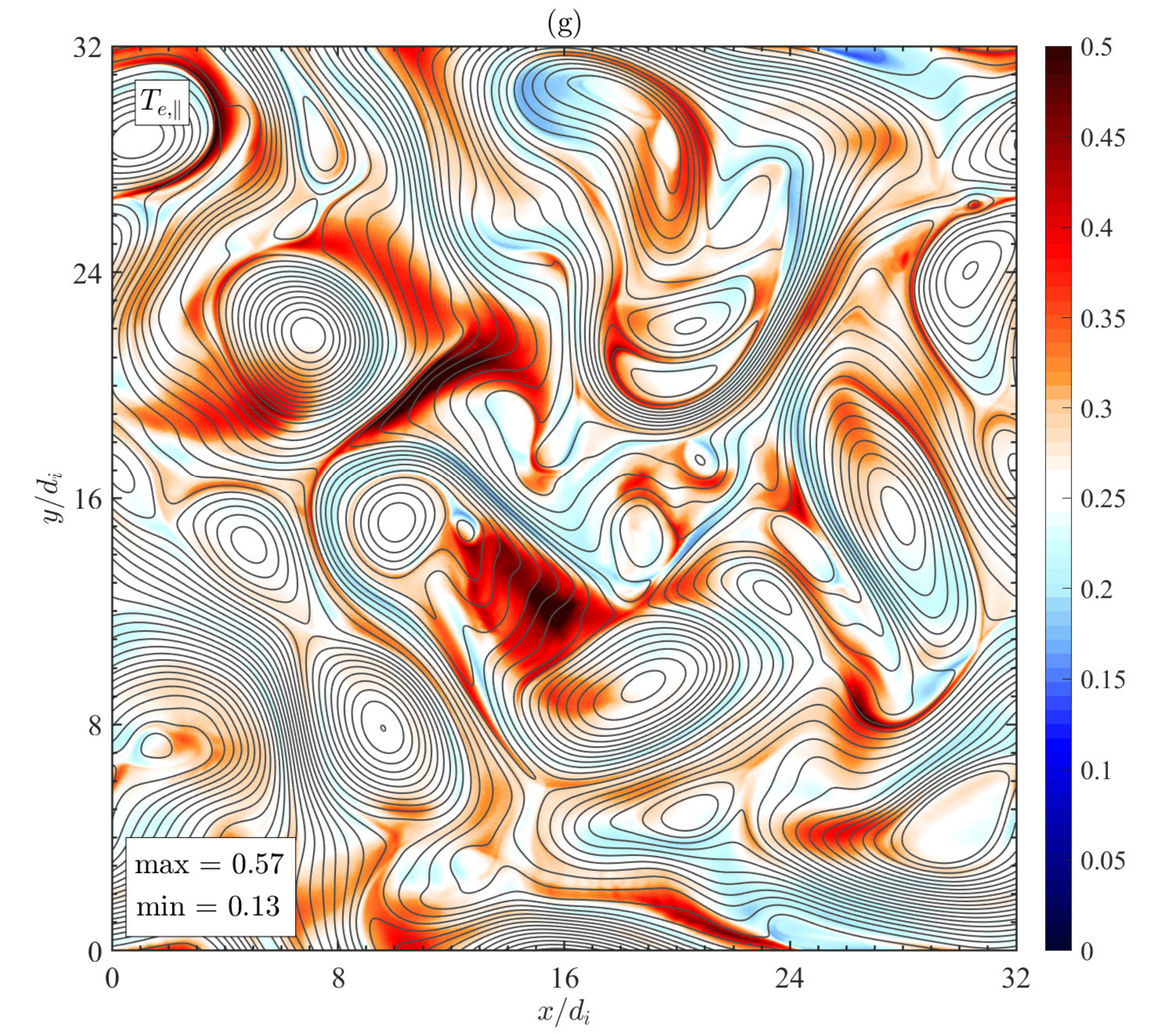}
\includegraphics[width=0.325\textwidth]{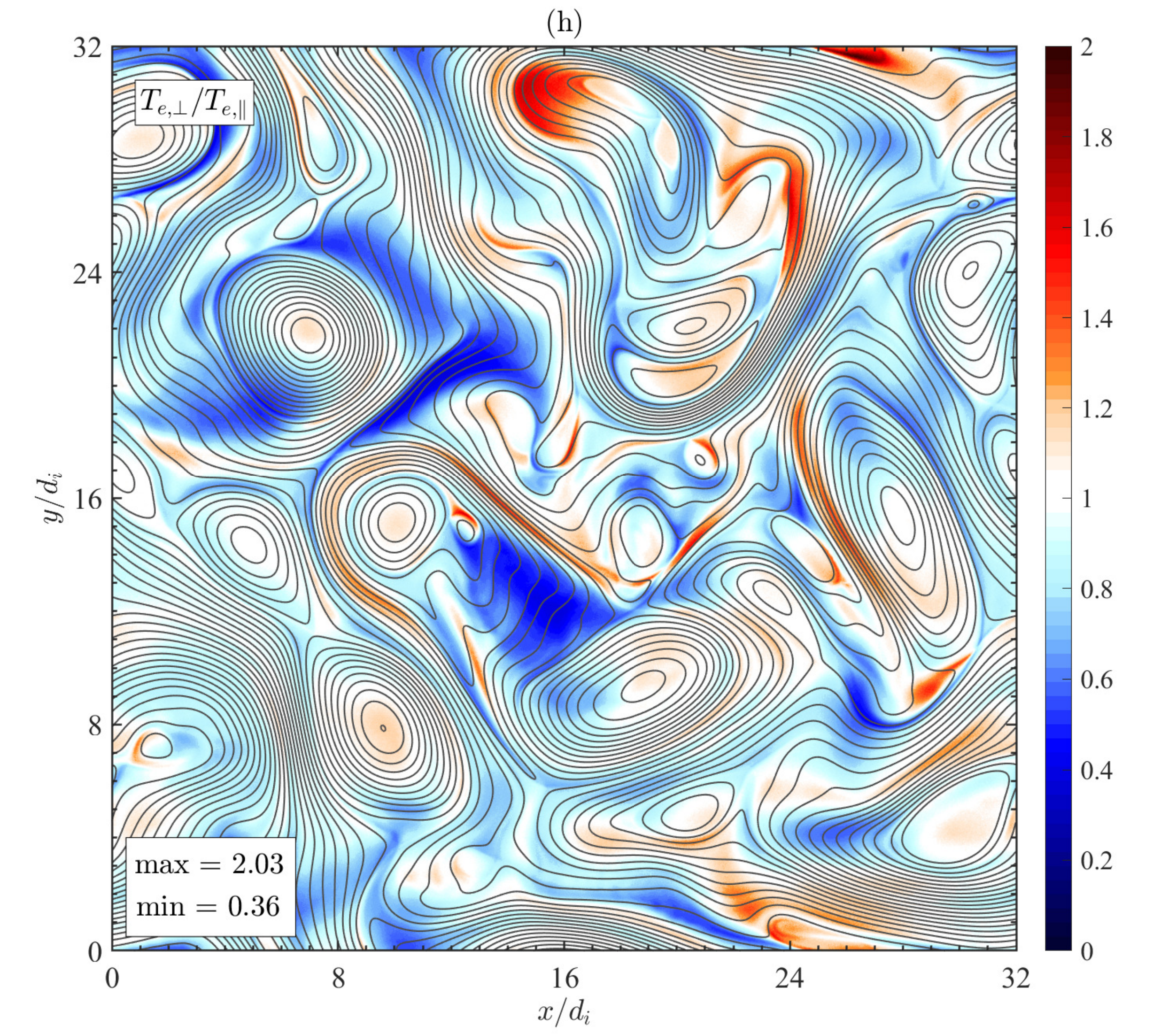}
\caption{Ion and electron temperatures. \textit{Panel (a)}: time evolution of the (simulation-box) averaged perpendicular, $T_{i,\perp}$, parallel, $T_{i,\parallel}$, total temperature, $T_{i}$, and temperature anisotropy, $A_i$, for the ions. The latter is normalized to its initial value. The components are defined with respect to the \textit{local} mean magnetic field. \textit{Panel (b)}: same quantities as in panel (a) but for the electrons. \textit{Panels (c-e)}: contour plots of $T_{i,\perp}$, $T_{i,\parallel}$, and $A_i$. \textit{Panels (f-h)}: contour plots of $T_{e,\perp}$, $T_{e,\parallel}$, and $A_e$. In panels (c)-(h), the maximum and minimum values of each quantity are reported in the bottom left corner.}
\label{fig:temperatures}
\end{figure*}

We now focus on the electron and proton heating generated by turbulence. In Fig.~\ref{fig:temperatures}, we report the overall evolution of both ion and electron temperatures and their spatial distribution once turbulence is fully developed. Fig.~\ref{fig:temperatures}(a) shows the time evolution of the average value (over the whole simulation box) of the perpendicular, $T_{i,\perp}$, parallel, $T_{i,\parallel}$, and total temperature, $T_{i} = (2 \, T_{i,\perp} + T_{i,\parallel})/3$, and of the temperature anisotropy, $A_i = T_{i,\perp}/T_{i,\parallel}$, all normalized to their initial values. We recall here that the temperature components are defined with respect to the \textit{local} mean magnetic field. At the time of maximum turbulent activity, $T_i$ has increased of almost 8\% with respect to its initial value $T_{i}^{\textrm{in}} = 0.1$. Such increase does not stop at $t^{\mathrm{peak}}$, being linear in time afterwards. This hints at the possibility that, in the presence of an energy injection mechanism that keeps feeding the turbulent cascade and maintains it in a quasi-steady state over a longer time, the ion temperature could increase significantly more. $T_{i,\perp}$ starts increasing as soon as the simulation begins and keeps increasing throughout the whole evolution, with some oscillations until $t \simeq 5$ and then a linear increase in time. $T_{i,\parallel}$, instead, remains constant until $t \simeq 2$ (comparable to the eddy turnover time at the injection scale), then decreases reaching a minimum just before $t^{\mathrm{rec}}$ and it increases monotonically afterwards. Correspondingly, $A_i$ increases until $t^{\mathrm{rec}}$ and then it decreases very slowly, reaching an almost constant value of $1.06$ at $t^{\mathrm{peak}}$.
The corresponding quantities for the electrons are shown in Fig.~\ref{fig:temperatures}(b). $T_{e}$ remains almost constant until about $t^{\mathrm{rec}}$, then increases very slowly, reaching a value of $1.01 \, T_{e}^{\textrm{in}}$ at $t^{\textrm{peak}}$, with $T_{e}^{\textrm{in}} = 0.25$ being its initial value. Despite the quite small increase in the total electron temperature, its perpendicular and parallel components exhibit very large variations. $T_{e,\perp}$ shows a small initial decrease, then stays almost constant until just before $t^{\mathrm{rec}}$ and then starts decreasing more rapidly, linearly in time. On the contrary, $T_{e,\parallel}$ increases initially more or less linearly in time, then at around $t^{\mathrm{rec}}$ it starts increasing much faster. Correspondingly, $A_e$ decreases monotonically throughout the simulation (faster after $t^{\mathrm{rec}}$), reaching a value of about $0.9$ at $t^{\textrm{peak}}$, which keeps decreasing afterwards. 

Summarizing, Fig.~\ref{fig:temperatures}(a)-(b) show that at $t \simeq t^{\mathrm{rec}} = 4.9 \, \Omega_i^{-1}$ (i.e., when the first maximum of $|\vect{J}|$ marks the onset of magnetic reconnection), there is a clear and significant change of behavior in the particle heating, characterized by the fact that: (i) the rate of variation of the total ion temperature increases and becomes almost constant; (ii) the total electron temperature starts increasing linearly in time; (iii) the ion temperature anisotropy reaches a plateau and starts decreasing very slowly; (iv) the rate of variation of the electron temperature anisotropy increases. It is then reasonable to assume that the change in particle heating is related to the onset of magnetic reconnection events.

In order to test this assumption, Fig.~\ref{fig:temperatures}(c)-(h) show the spatial distribution of the above quantities in the 2D simulation domain. In the top row (panels \textit{c-e}), we show the ion perpendicular and parallel temperatures and their ratio. $T_{i,\perp}$ and $T_{i,\parallel}$ exhibit a very patchy behaviour, with alternating regions where they are larger or smaller of their initial value. Typically, in regions where one component has increased the other one has decreased. As a result, despite their average values are very similar and their maximum values are also comparable, $A_i$ is larger than 2 in many large areas, reaching a maximum as large as 3.7. Regions with values well above or below 1 are alternating and somewhere we observe a quadrupolar configuration, both in correspondence of X-points where magnetic lines reconnect (e.g., at the point [8,17]) and inside big vortexes (e.g., at [19,9]).  
The spatial distribution of the electron temperature looks quite different. $T_{e,\perp}$ is close to its initial value in most of the simulation box, exhibiting just a small decrease in correspondence of what look like reconnection outflows around X-points and a small increase at the center of vortexes. On the contrary, $T_{e,\parallel}$ exhibits much larger variations, slightly decreasing in regions where the magnetic field lines get compressed and becoming about twice larger in the reconnection outflows. As a consequence, in these latter areas $A_e$ gets very small reaching values as small as $0.36$.
The spatial distribution of both ion and electron temperatures is consistent with what we have inferred from the time evolution of their average values and with the observed change of behavior at the time when reconnection starts occurring: reconnection is likely to be at least partially responsible for particle heating, especially for the electrons. This sounds reasonable, as magnetic reconnection is a mechanism that transfers energy from the magnetic field to the particles.

\subsection{Standard and electron-only reconnection}
\label{subsec:reconnection}

\begin{figure*}
\centering
\includegraphics[width=0.32\textwidth]{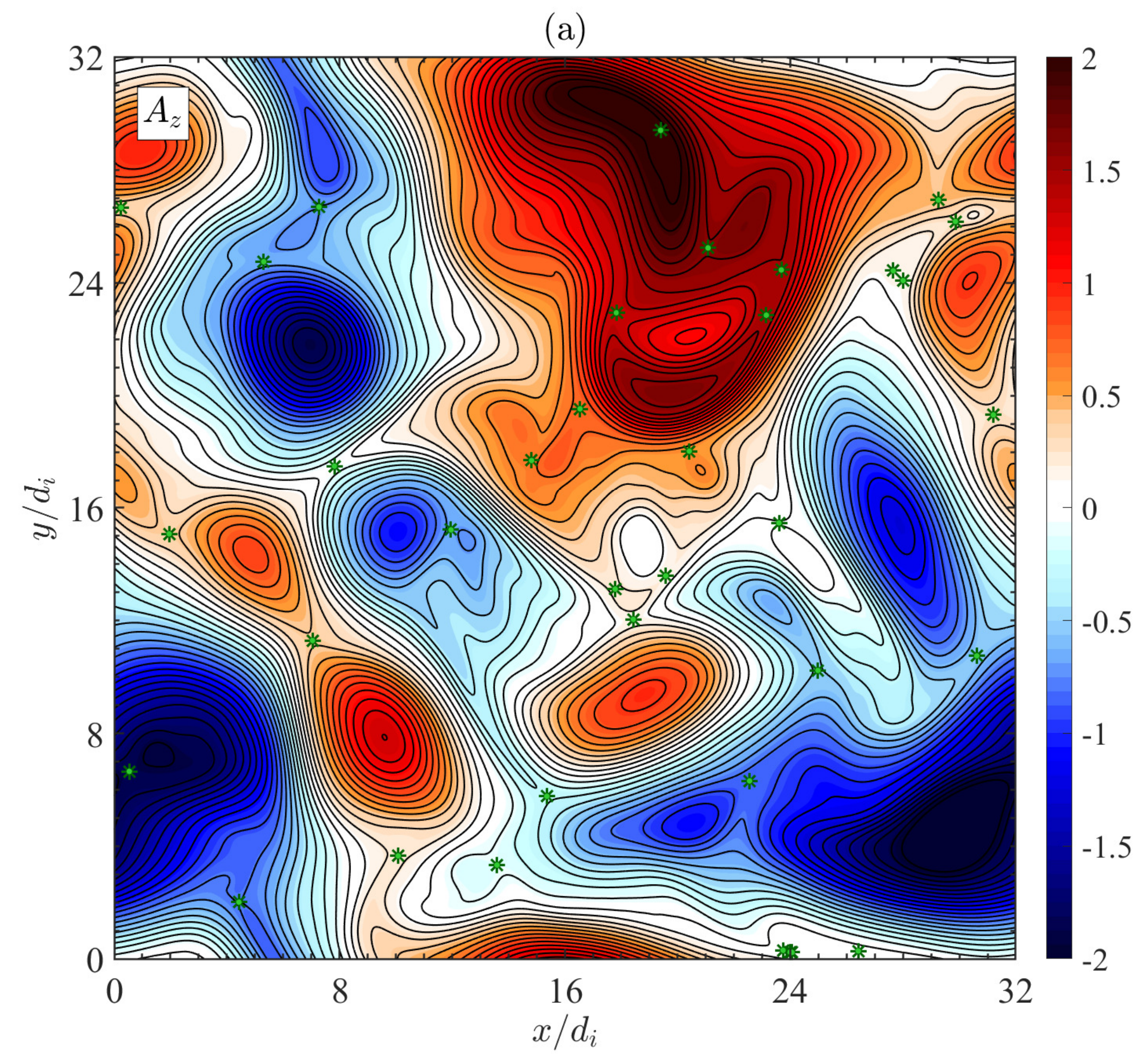}
\includegraphics[width=0.32\textwidth]{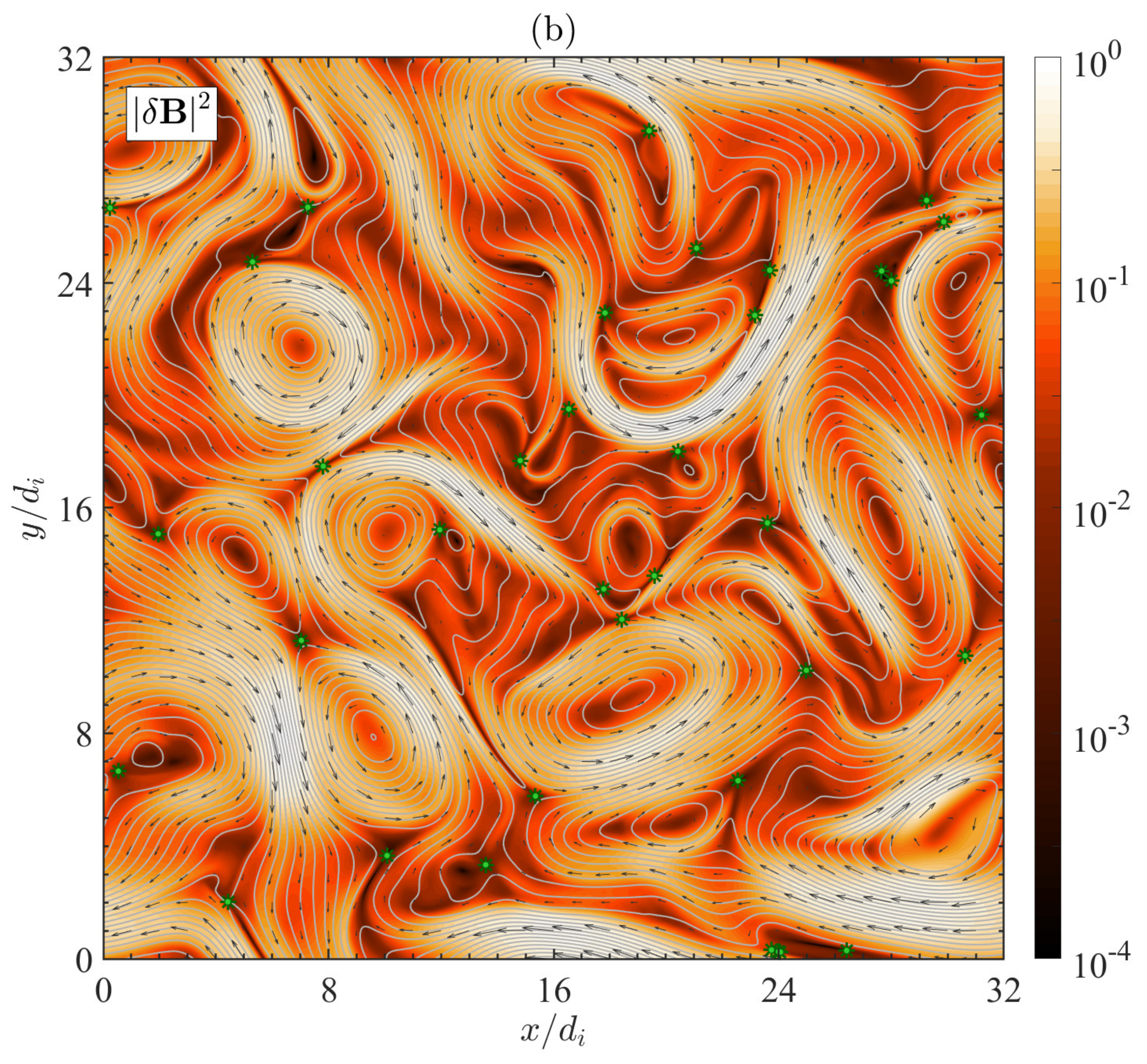}
\includegraphics[width=0.32\textwidth]{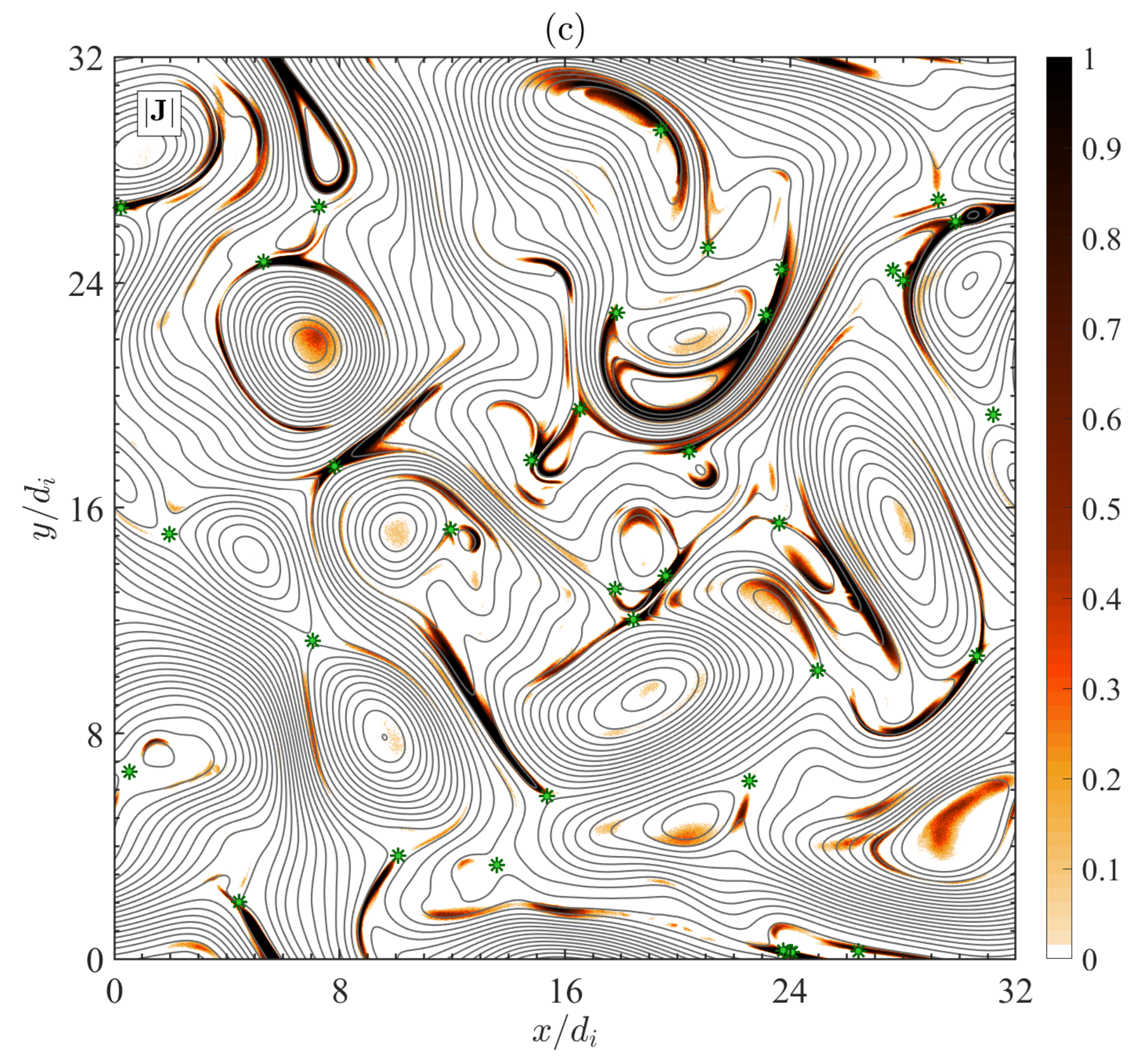}\\
\includegraphics[width=0.32\textwidth]{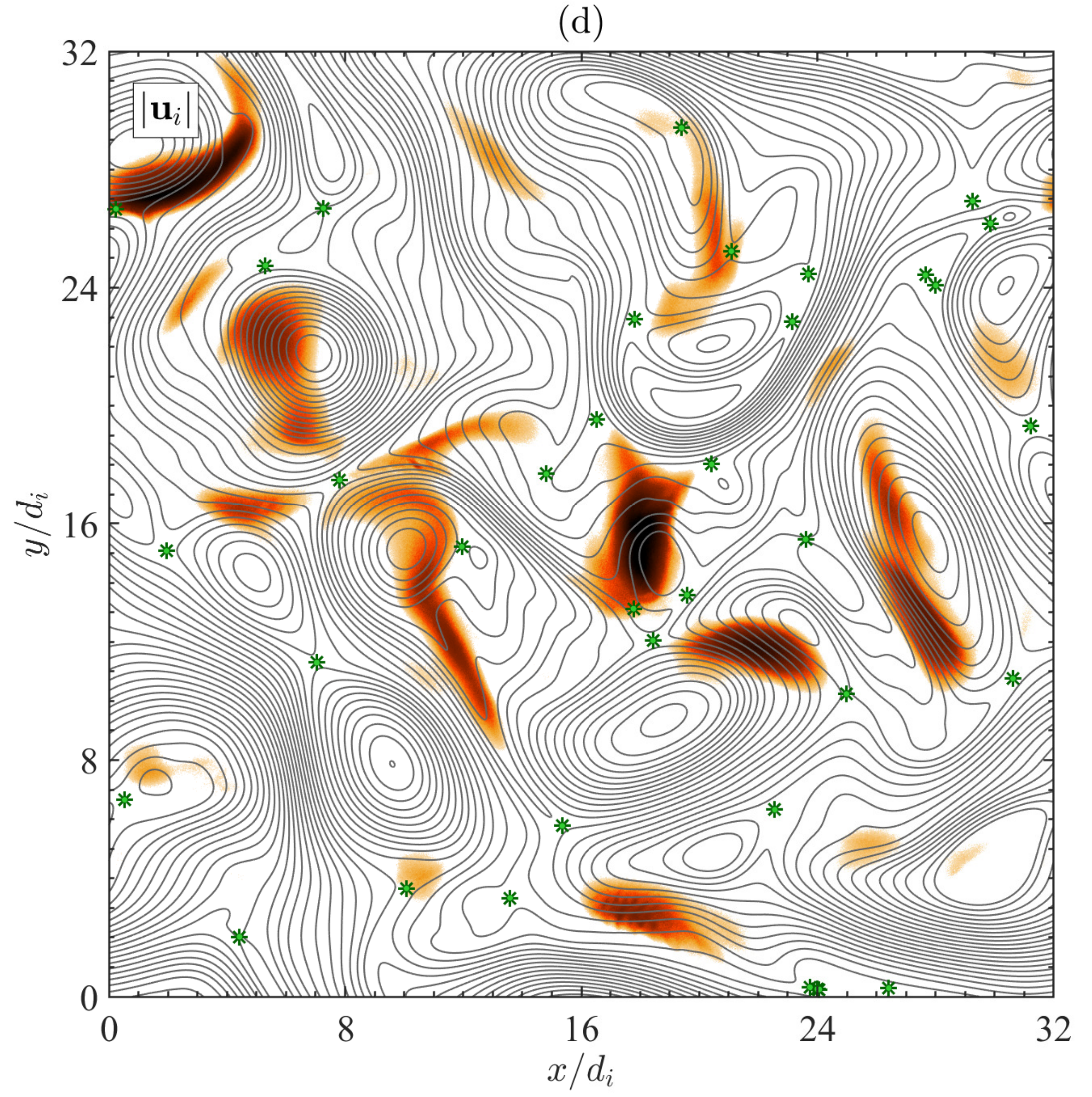}
\includegraphics[width=0.32\textwidth]{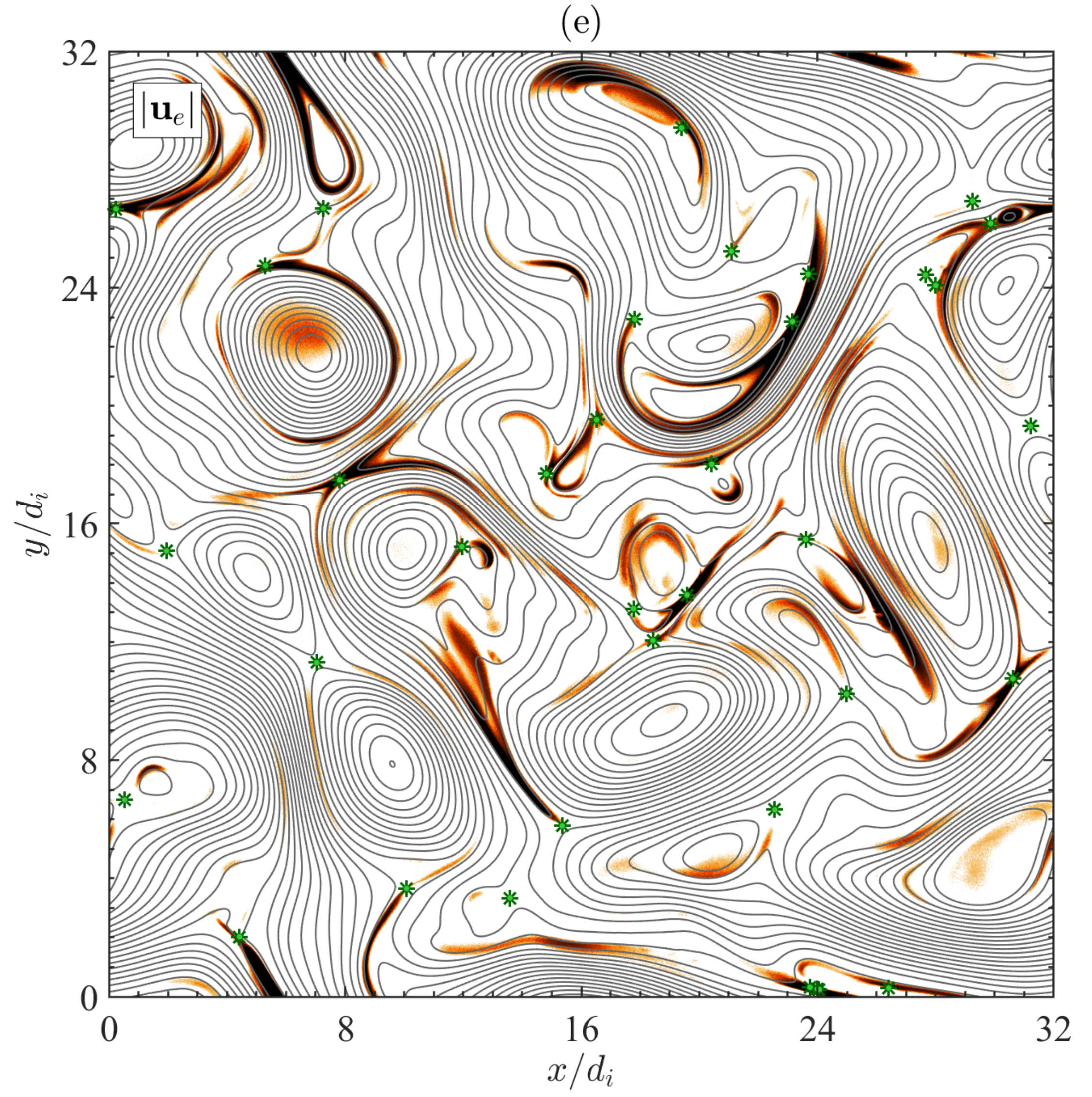}
\includegraphics[width=0.32\textwidth]{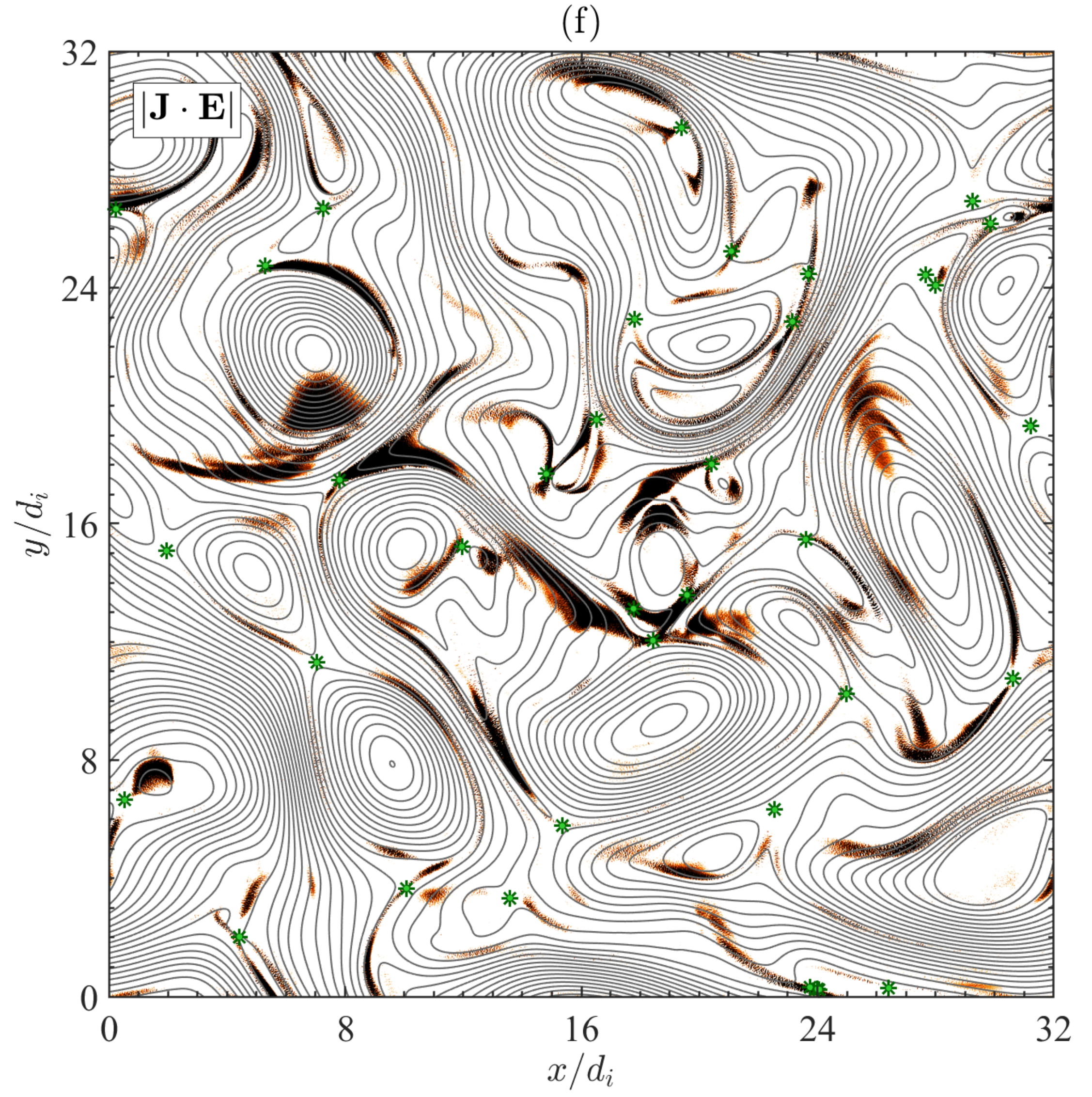}\\
\includegraphics[width=0.32\textwidth]{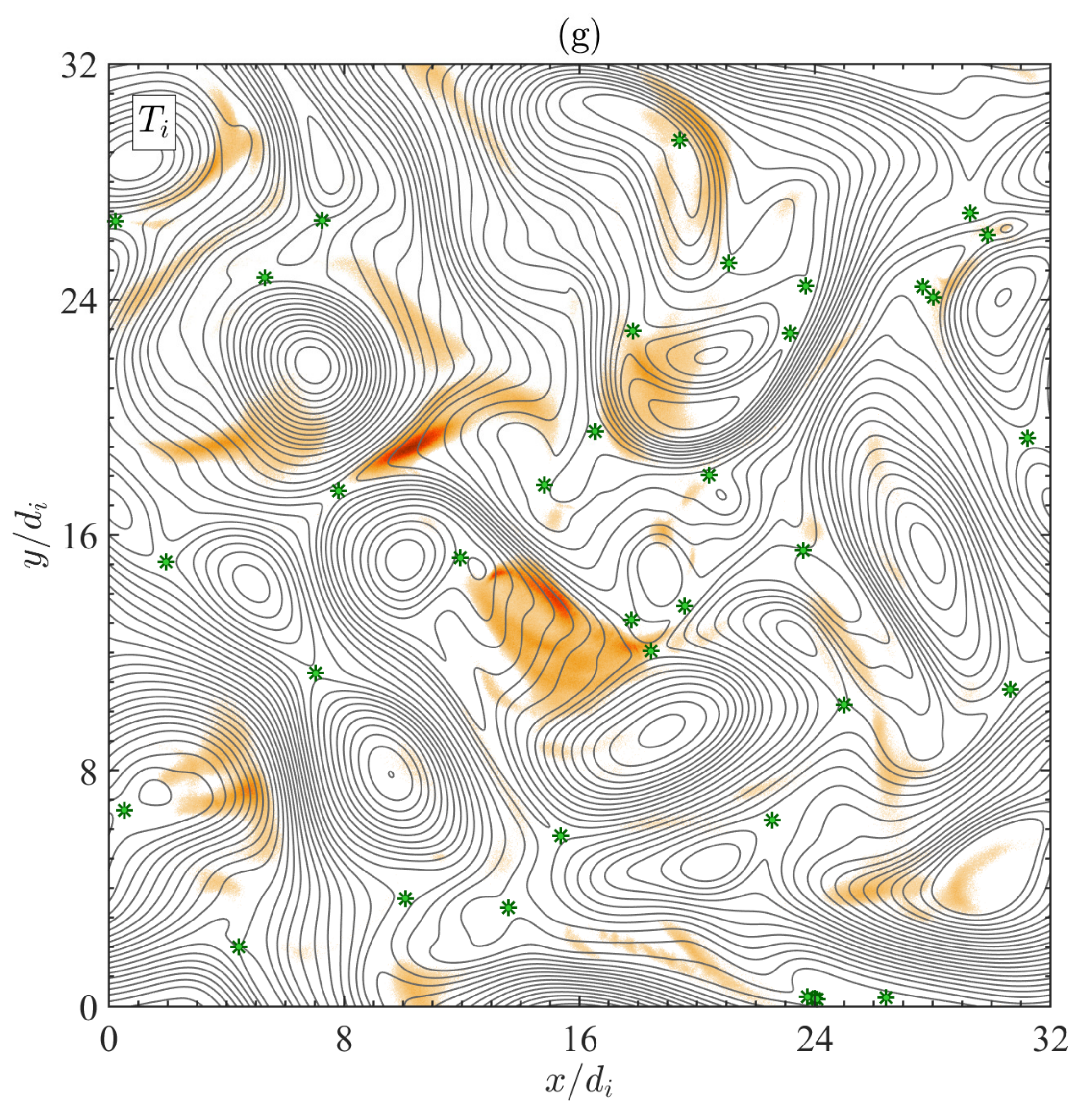}
\includegraphics[width=0.32\textwidth]{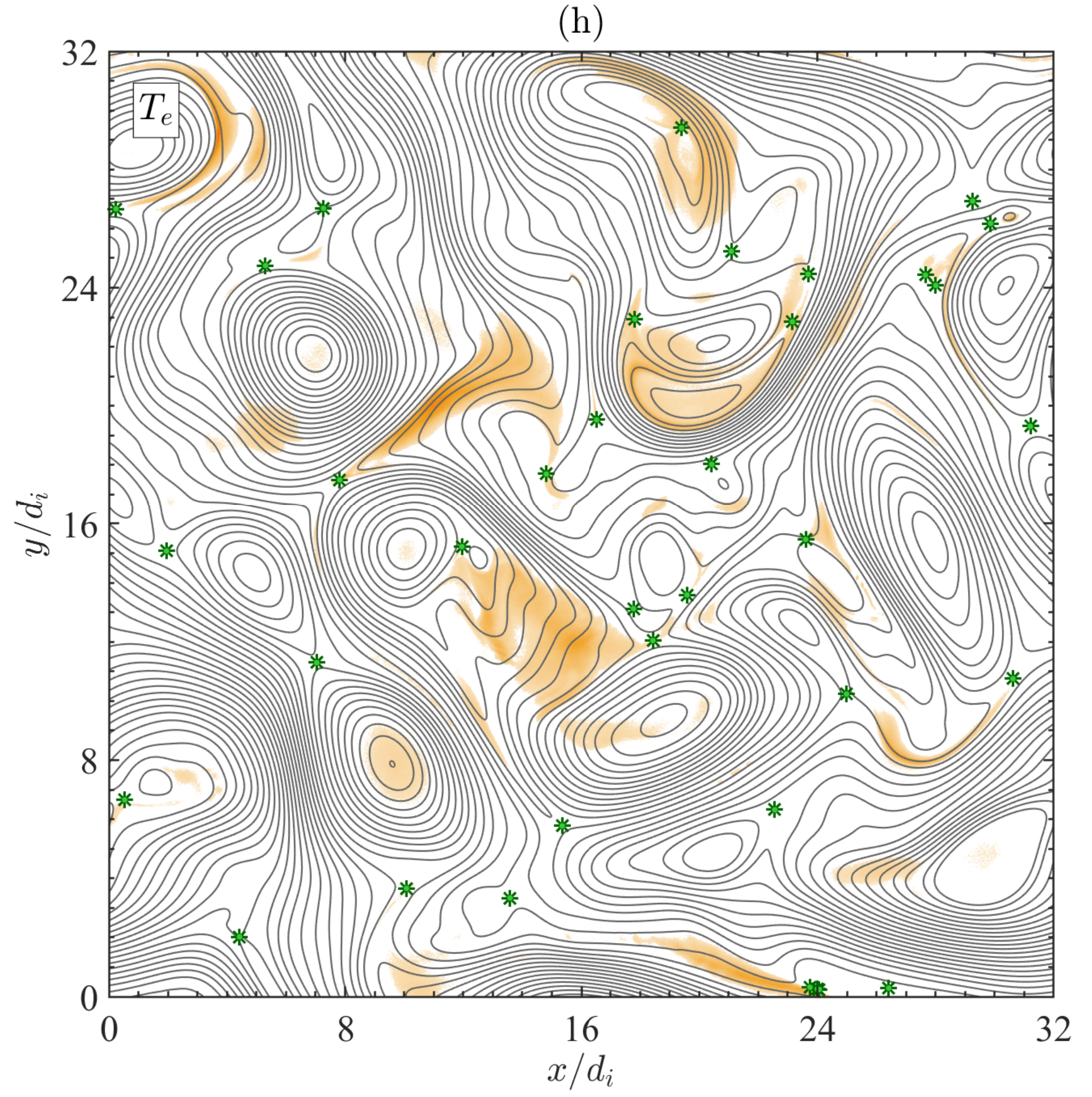}
\includegraphics[width=0.32\textwidth]{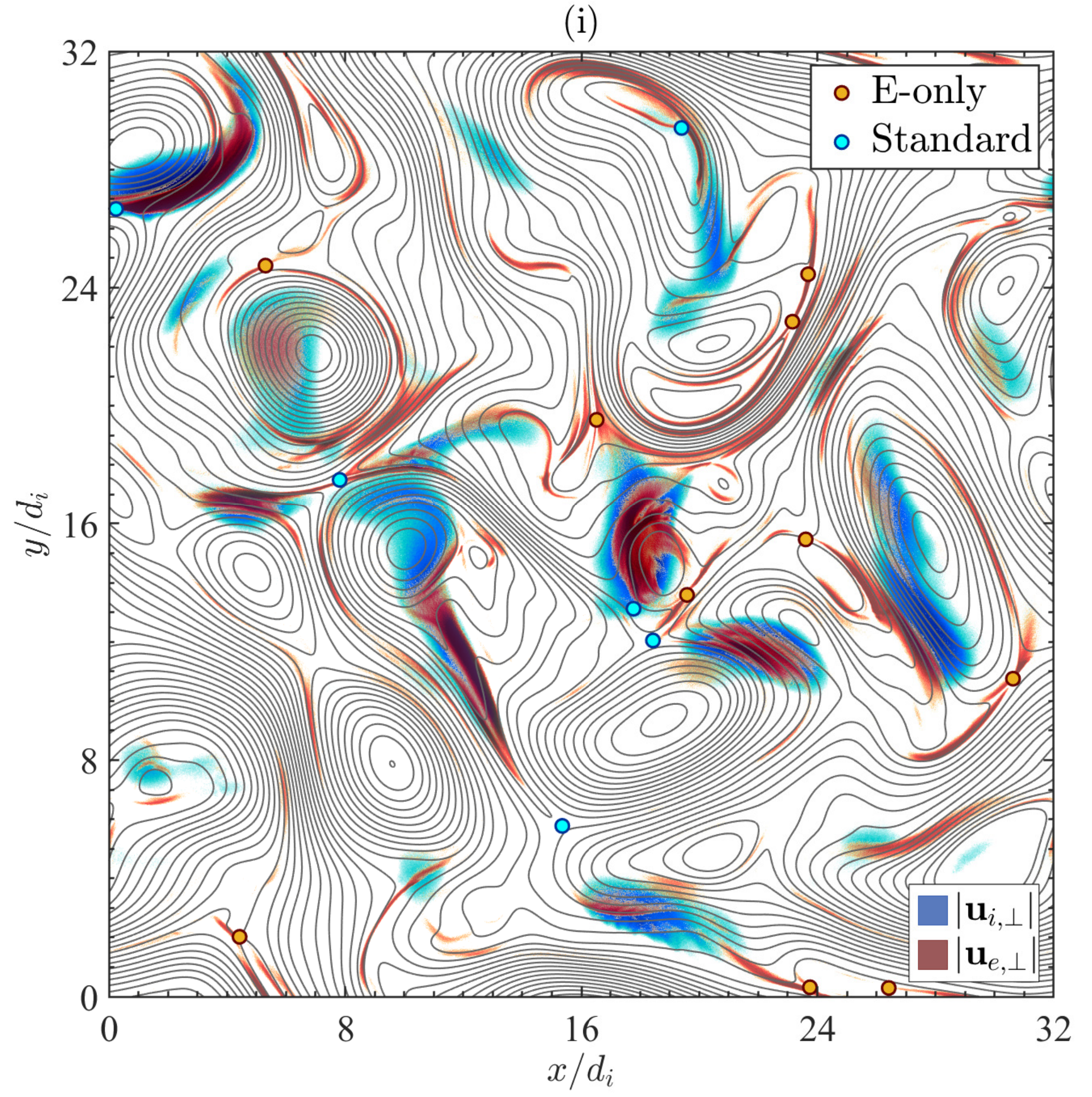}\\
\caption{Identification of magnetic reconnection events through the indicators C1-C4 from~\citet{Agudelo_al_2021}. \textit{Panel (a)}: contour plot of the out-of-plane vector potential $A_z$. The black isocontours represent magnetic field lines. Green stars mark the saddle points, which represent potential reconnection sites. \textit{Panel (b)}: contour plot of the magnitude of the magnetic fluctuations, with arrows marking the direction of the local magnetic field. \textit{Panel (c-h)}: contour plots of $\Psi/\Psi^{\mathrm{th}}-1$, with $\Psi = |\vect{u}_{i}|$, $|\vect{u}_{e}|$, $|\vect{J} \cdot \vect{E}|$, $T_{i}$, and $T_{e}$, respectively. These panels have all the same color scale as in panel (c). \textit{Panel (h)}: classification of some reconnection events as ``standard'' or ``electron-only'' by superimposing the areas where $|\vect{u}_{e,\perp}|$ (red) and $|\vect{u}_{i,\perp}|$ (blue) exceed their respective thresholds.}
\label{fig:reconnection}
\end{figure*}

In order to confirm the role of reconnection in heating the particles and in generating a strong temperature anisotropy, especially for the electrons, we now provide a more quantitative characterization of the regions where reconnection events occur.
Figure~\ref{fig:dB2} clearly showed qualitative evidence for the presence of reconnection events in the simulation, e.g., X-points and small magnetic islands emerging in proximity of thin strong current sheets. Magnetic reconnection is known to occur in turbulent plasmas as the the result of the interaction of turbulent eddies which leads to the formation and stretching of intense current sheets. Indeed, it has been observed to occur spontaneously in numerical simulations of plasma turbulence performed with different methods~\citep[e.g.][]{Matthaeus_1986,Servidio_al_2009,Franci_al_2015a,Wan_al_2015,Cerri_Califano_2016,Haggerty_al_2017} and it has been shown to provide a significant contribution to the further development of the turbulent cascade at sub-ion scales~\citep[e.g.][]{Franci_al_2017,Mallet_al_2017b,Papini_al_2019}. In Fig.~\ref{fig:reconnection}(a), we show a pseudocolor plot of the out-of-plane component of the vector potential, $\vect{A}_z$, with its isocontours superimposed as black lines. Green stars mark saddle points, i.e., X-points which represent potential reconnection sites. 
Fig.~\ref{fig:reconnection}(b) shows arrows marking
the direction of the local magnetic field in the simulation box, on top of the contour plot of the $|\delta \vect{B}|^2$. This clearly shows that the local magnetic field has opposite sign at the two sides of most of the X-points, as required for magnetic reconnection to occur.
Recently, \citet{Agudelo_al_2021} have investigated the spontaneous occurrence of reconnection in a 3D fully kinetic simulation of plasma turbulence by means of a set of indicators based on thresholds. Here we chose to apply the same criteria to detect and characterize reconnection events. Although our 2D setting is definitely less realistic than a 3D one, here we have the advantage that it is straightforward to identify X-points in 2D. In this sense, our analysis of reconnection events can be considered as complementary to the work by \citet{Agudelo_al_2021}, and in some sense a further validation.
For a given (positive) scalar quantity, $\Psi$, we define a threshold $\Psi_{\textrm{th}} = \langle \Psi \rangle + \mathcal{N} \Psi^{\textrm{rms}}$ and we look for regions where  $\Psi/\Psi_{\textrm{th}}-1 > 0$. Here, we chose to set $\mathcal{N} = 1$, as we find that this is enough for detecting the strongest reconnection events in our simulation.
The  criteria we have evaluated are:
\begin{alignat*}{2}
\mathrm{C1}&: |\vect{J}| > |\vect{J}|^{\mathrm{th}} \equiv \langle |\vect{J}| \rangle + |\vect{J}|^{\textrm{rms}} \;\; & \mathrm{(Fig.~}\ref{fig:reconnection}\mathrm{c)}\\
\mathrm{C2}&:  |\vect{u}_{i,e}| > |\vect{u_{i,e}}|^{\mathrm{th}}  \equiv \langle |\vect{u}_{i,e}| \rangle + |\vect{u}_{i,e}|^{\textrm{rms}} \;\; & \mathrm{(Fig.~}\ref{fig:reconnection}\mathrm{d,e)}\\
\mathrm{C3}&: T_{i,e} > T_{i,e}^{\mathrm{th}} \equiv \langle T_{i,e} \rangle + T_{i,e}^{\textrm{rms}} \;\; & \mathrm{(Fig.~}\ref{fig:reconnection}\mathrm{g,h)}\\
\mathrm{C4}&: \vect{J} \cdot \vect{E}|_{+} > \vect{J} \cdot \vect{E}|_{+}^{\mathrm{th}} \equiv \langle \vect{J} \cdot \vect{E}|_{+} \rangle + \vect{J} \cdot \vect{E}|_{+}^{\textrm{rms}} \;\; & \mathrm{(Fig.~}\ref{fig:reconnection}\mathrm{f)}
\end{alignat*}
where for the last criterion, we consider only positive values of $\vect{J} \cdot \vect{E}$, which denote energy transfer from the fields to the particles.
Figure~\ref{fig:reconnection}(c) shows the result of indicator C1 for $|\vect{J}|$, revealing the presence of intense current structures. We can clearly see that all these have a width of the order of $d_e$ and lengths of the order of a few times $d_i$. We count about a dozen of them, taking into account that some of them are actually portions of the same current sheet, as the simulation box is periodic. In Fig.~\ref{fig:reconnection}(d)-(e), we estimate the indicator C2 for $|\vect{u}_{i,e}|,$ related to the presence of fast ions and electrons that provide evidence for the presence of a  reconnection outflow (or jet). The ion bulk velocity exceeds the threshold in some regions with a width of the order of $1-2 \, d_i$, some of which seem to be directly related to X-points and represent ion jets in the reconnection outflow, while others seem to be within large-scale vortexes. On the contrary, the electron bulk velocity is above its threshold in all the regions directly connected to the strong current sheets observed in panel (c). These fast electron jets have a very small width, of the order of the $d_e$, as they are confined in small regions between larger-scale structures. It is interesting to note that many of the observed reconnection events seem to be highly asymmetric, as jets are observed only on one side of the X-point and not on the other. Figure~\ref{fig:reconnection}(f), shows the contour plot of indicator C4. $\vect{J} \cdot \vect{E}|_{+}$ is above its threshold in a few thin regions corresponding to the most intense current sheets, providing further evidence that in the location of reconnection events magnetic energy is converted into particle energy. In Fig.~\ref{fig:reconnection}(g)-(h), we estimate indicator C3 on $T_{i,e}$ which is related to the presence of heated ions and electrons. The regions where the particle temperatures exceed their respective threshold look similar for ions and electrons, meaning that most reconnection events are eventually leading to both ion and electron heating. A major difference, however, can be appreciated by comparing with Fig.\ref{fig:temperatures}(c)-(h): reconnection heats the electrons preferentially in the parallel direction with respect to the local magnetic field and the ions in the perpendicular direction. The latter result appears to be an agreement with the fact that a sheared in-plane ion flow tends to transfer energy to the in-plane ion pressure tensor components via the action of the symmetric part of the strain tensor \citep{DelSarto_al_2016,Yang_al_2017,DelSarto_Pegoraro_2018,Matthaeus_al_2020,Bandyopadhyay_al_2021,Hellinger_al_2022}. Again, we observe a significant lack of symmetry, as particles are heated only on one side of the X-point, typically the same for both species. 
In Fig.\ref{fig:reconnection}(i), we compare the indicators C2 for ions and electrons, by showing the contour plots of their perpendicular components, $|\vect{u}_{i,\bot}|^2$ and $|\vect{u}_{e,\bot}|^2$ on top of each other (in shades of blue and red, respectively). This allows us to appreciate if a reconnection event produces both in-plane ion and electron jets, as in ``standard'' reconnection, or only an electron jet with no ion counterpart, as in ``electron-only'' reconnection. For some of the X-points, it is not straightforward to apply this classification, mainly because no strong ion nor electron jets are observed or because the geometry is complex. We still observe about a dozen of events which can be labelled as electron-only (yellow circles) and about half a dozen that are standard (light blue circles).
Summarizing Fig.~\ref{fig:reconnection}, we can conclude that as a result of the interaction between turbulent magnetic structures, thin intense current sheets form and shrink until they undergo reconnection. Such reconnection events transfer magnetic energy to the particles, heating the electrons in the direction parallel to the mean field and the ions mainly in the orthogonal direction. Thin electrons jets are also observed in the reconnection outflow, with ion counterparts only in about one third of events. The fact that standard reconnection is observed only in a minority of events might be related the simulation box size, which is only a few tens of $d_i$, and to the energy injection, which occurs at the ion scales~\citep{Califano_al_2020}. These conditions strongly limit the magnetic correlation length, which characterizes the size of the interacting turbulent magnetic structures and, as consequence, also set the length and thickness of the current sheets forming in-between~\citep[e.g.][]{Stawarz_al_2019}. In other words, there are only a few vortexes with radius of the order of few times $d_i$, so there are only a few chances to develop strong current sheets with such a length. On the contrary, most of them are formed by the interaction of magnetic structures with smaller size and will therefore be shorter. Indeed, the ion jets observed in Fig.~\ref{fig:reconnection}(i) seem to be concentrated just next to some of the largest vortexes.
Quantitatively characterizing the properties of reconnection and its interplay with electron-scale turbulence goes beyond the scope of this work and will be the subject of future investigation. Here, we intended to show evidence for the coexistence of both types of reconnection events spontaneously driven by turbulence at sub-ion scales and at electron scales. 

\section{Discussion and conclusions}
\label{sec:conclusions}
In this work, we have presented the results of a two-dimensional simulation of plasma turbulence at sub-ion and electrons scales performed with the fully kinetic code iPIC3D, showing signature of turbulence-driven magnetic reconnection and related anisotropic particle heating. We have modelled plasma conditions similar to those measured by PSP during its first solar encounter. This allowed us to  extend towards larger wavenumbers the analysis of the ion-scale turbulent cascade performed with hybrid simulations in \citet{Franci_al_2020arxiv}.
Our simulation employs a large box in terms of the electron characteristic scales, i.e., $320 \, d_e \times 320 \, d_e$ ($d_i = 10 \, d_e$, since the proton-to-electron mass ratio has been set to $100$), a spatial resolution $\Delta x = d_i/64 \simeq 0.16 \,d_e$, and implements a very large number of particles (1024 ppc ions and 8192 ppc electrons, for a total of more than 40 billion particles in the whole simulation domain). This setting allowed us to accurately model  the development of the turbulent cascade and to determine the spectral properties of the magnetic and electron bulk velocity fluctuations for a decade and a half in wavenumber in the range $0.1 \lesssim k_\bot d_e \lesssim 4$, fully capturing the electron-scale transition.


The turbulent dynamics is characterized by the concurrent presence of coherent magnetic field structures in the form of vortexes, with radii between a few times $d_e$ and a few times $d_i$, and very elongated thick filaments, with a width of the order of $d_i$ and a length up to about about $20 \,d_i$. In between these, thin intense current sheets form, with width of the order of $d_e$.

Our results show that, at the maximum of turbulent activity, the power spectrum of the magnetic fluctuations exhibits is very well modelled by a double power law, with a spectral index $\alpha_{B}$ compatible with $-11/3$ and $-5$ respectively above and below the electron characteristic scales. Since the electron beta is close to $1$, the electron inertial length and gyroradius are very close to each other, thus making it impossible to infer whether the electron-scale transition is associated to either or both of them. Complementary numerical simulations (not shown here) performed by varying both spatial resolution and number of ppc, have showed that the location of such transition is of physical origin, provided that a sufficient number of grid points allows to cover approximately a full decade across the electron scales. The power spectrum of the electron bulk velocity behaves just as one would expect considering that at sub-ion scales the current density is almost exclusively due to the electron bulk motion: it also behaves like a double power law, with a spectral index $\alpha_{e} = \alpha_{B}+2$.  
We observe a hint of a further steepening in the power spectrum of the electron bulk velocity fluctuations (and correspondingly, of the magnetic field) at $k_\bot d_e \gtrsim 2$. However, such steepening occurs very close to the scale at which the spectra reach the noise level, therefore it is not possible to draw any conclusion. Simulations with a better spatial resolution will be needed to further investigate whether such steepening is present or not. 
The spectrum of the electric fluctuations behaves differently, as it does not exhibit any major change of slope when the electron scales are reached. Such behavior seems to be consistent with Cluster observations in the Earth's magnetosheath~\citep{Matteini_al_2017}, where the steepening of the spectrum of the magnetic field at electron scales does not have a counterpart in the electric field.
A thorough investigation of the nature and of the spectral properties of the electric field, including an analysis of the different contributions to the generalized Ohm's law, is currently ongoing and will be the subject of a follow-up work.

The limited size of the simulation box in terms of $d_i$, together with the fact that the initial energy injection in the simulation occurs at the ion scales, makes it impossible to model the turbulent cascade in the inertial MHD range. We have compared the power spectra of all the fields with those obtained from a hybrid kinetic simulation (performed with the CAMELIA code) which implemented the same plasma conditions (same ion beta, same eletron beta, very similar initial amplitude of the turbulent fluctuations) but used a much larger box. Indeed, all power spectra from the fully-kinetic simulation are well in agreement with their hybrid counterparts, down to scales comparable or slightly larger than $d_e$. Below this scale, almost all the spectra from the iPic3D simulation further steepen, while the ones from CAMELIA exhibit no change of behavior. This is not surprising, as the two models differ significantly at the electron scales: CAMELIA treats the electrons as an isothermal massless fluid and therefore electron inertia effects and electrons kinetic effects are not retained. What is not necessarily obvious is the fact that the spectral behavior below $d_i$ is almost completely unaffected by the existence/absence of a turbulence cascade in the inertial range. This hints at a certain degree of universality of the sub-ion scale turbulent cascade: its properties seem to depend on the plasma conditions alone, regardless of what is happening at larger scales. This is in agreement with what has been recently observed in~\citet{Franci_al_2020}. There, the spectral properties of turbulence driven by a Kelvin-Helmholtz instability observed by MMS in the Earth's magnetosheath have been correctly modelled by a hybrid simulation of Alfv\'enic turbulence. This employed the observed values for the plasma beta, the electron beta, and the level of turbulent fluctuations with respect to the ambient magnetic field. 

The development of turbulence is observed to have an effect on particle heating. The ions quickly develop a certain temperature anisotropy, whose average value reaches a maximum of $1.08$ at $t \simeq t^{\mathrm{rec}}$ and then decreases very slightly. Their average parallel temperature with respect to the local magnetic field starts increasing at $t \simeq t^{\mathrm{rec}}$ and reaches the average perpendicular component at $t \simeq t^{\mathrm{peak}}$. This behavior is quite similar to what observed in previous hybrid simulations, where the perpendicular ion temperature quickly starts to increase, leading to the formation of temperature anisotropy, with a comparable increase of the parallel temperature starting later (more or less at time of the the onset of magnetic reconnection), therefore freezing the temperature anisotropy ~\citep{Franci_al_2015a,Franci_al_2017}.
The electrons develop a temperature anisotropy in the opposite sense, i.e., the parallel component is much larger than the perpendicular and whose average value keeps decreasing throughout the whole evolution, even after the turbulent cascade has fully developed. Both the ions and the electrons are heated during the simulation, as their total temperature increases of $7 \%$ and $1 \%$, respectively. In the time evolution of the temperature of both species we observe a link with the onset of magnetic reconnection: at $t \simeq t^{\mathrm{rec}}$, the rate of increase of the ion temperature sharply increases, while the electron temperature starts increasing after remaining at its initial value in the first part of the simulation.

The spatial distribution of the electron temperature and its anisotropy confirms a major role of magnetic reconnection in heating the electrons: the regions where the parallel electron temperature increases (and therefore where the temperature anisotropy mostly differs from $1$) are localized in the outflows of reconnection events at the sides of X-points. 
This has been confirmed by detecting reconnection events using the criteria defined and applied in~\citet{Agudelo_al_2021} (there for a 3D fully kinetic simulation), which are based on intensity threshold on the current density, the ion and electron bulk velocity and temperature, and the energy transfer from the electromagnetic fields to the particles mediated by $\vect{J} \cdot \vect{E}$. The results of such analysis have revealed the coexistence of ``standard'' and ``electron-only'' reconnection events, exhibiting both ion and electron jets or electron jets alone, respectively. There is a hint that the different nature of reconnection events is related to the length of the reconnecting current sheet, which in turns depends on the size of the interacting magnetic vortexes~\citep{Stawarz_al_2019}. The latter is also directly link to the correlation length of the turbulence, which has been observed to have an impact on the nature of the reconnection~\citep{Califano_al_2020,SharmaPyakurel_al_2019,Stawarz_al_2022}. 
The areas where we observe the largest increases in electron temperature are all located in reconnection outflows. For the ions, some areas of increased temperature are in correspondence of reconnection events, while others appear to be inside magnetic field vortexes.

In conclusion, we have modelled the development of plasma turbulence accurately from ion down to sub-electron scales, observing the spontaneous occurrence of magnetic reconnection events (both standard and electron-only), which are associated with strongly anisotropic electron heating. 

In order to make the simulation computationally feasible, in this work we employ smaller values of some plasma parameters with respect to their real value or to their typical value in the solar wind at 1 AU: $m_i/m_e = 100$ instead of $1836$ (as a consequence, $d_i = 10 \, d_e$ instead of $43 \, d_e$), $c/v_{Ai} = \omega_i/\Omega_i = 200$ instead of $\sim 10^3$ and $c/v_{Ae} = \omega_e/\Omega_e = 20$ instead of $\sim 200$. This is consistent with the fact that here we are modelling the near-Sun solar wind, as encountered by PSP during its first perihelion. In this sense, our parameters are just intermediate between the typical solar-wind values mentioned above and those in the upper solar corona, i.e., $c/v_{Ai} \sim 10^2$ and $c/v_{Ae} \sim 3$. 
\citet{Verscharen_al_2020} demonstrated that plasma models employing $m_i/m_e = 100$ and
$c/v_{Ai} \gtrsim 10$ can successfully cover physics on scales $\gtrsim 0.2 \,d_i$ for $\beta_i \sim \beta_e \sim 1$, which is the regime we have explored here. Since we employ that exact value of $m_i/m_e$, an order of magnitude larger value of $c/v_{Ai}$, and we also have $c/v_{Ae} \gtrsim 10$, we are confident that our fully kinetic simulation provide quite a correct modelling of the plasma dynamics down to electron scales.    

That said, a main limitation of our study is the 2D geometry, which strongly constrains our simulation results, since both turbulence and magnetic reconnection are inherently 3D processes. Modelling the spectral behavior reliably from ion to sub-electron scales and estimating particle heating quantitatively, however, require a high accuracy (in terms of both grid size and number of ppc) which are difficult to reach in 3D at the moment. Therefore, high-resolution 2D fully kinetic simulations represent an optimal starting point for this kind of analysis. Still, all the results obtained in the present study will need to be validated by future 3D simulations with similar plasma conditions. Indeed, this will be the subject of future work.

\section*{Acknowledgements}

The authors acknowledge valuable discussions with Julia Stawarz, Jeffersson Agudelo Rueda, Daniel Verscharen, Lorenzo Matteini, Domenico Trotta,  Francesco Califano, Giuseppe Arr\`o, and with the members of the Solar Orbiter Science Working Groups.

L.F. and D.B. are supported by the UK Science and
Technology Facilities Council (STFC) grant ST/T00018X/1. A.M. thanks the Belgian Federal Science Policy Office (BELSPO) for the provision of financial support in the framework of the PRODEX Programme of the European Space Agency (ESA) under contract number 4000134474. G.L acknowledges funding from the KULeuven Bijzonder Onderzoeksfonds (BOF) under the C1 project TRACESpace, from the European Union’s project DEEP-SEA (grant agreement 955606 ) and from the NASA grant 80NSSC19K0841.

This work was performed using the DiRAC Data Intensive Service at Cambridge, which is operated by the University of Cambridge Research Computing as part of the Cambridge Service for Data Driven Discovery (CSD3), and the DiRAC Data Intensive service at Leicester (DIaL), operated by the University of Leicester IT Services, both of which form part of the STFC DiRAC HPC Facility (www.dirac.ac.uk). The DiRAC component of CSD3 was funded by BEIS capital funding via STFC capital grants ST/P002307/1 and ST/R002452/1 and STFC operations grant ST/R00689X/1. The DIaL equipment was funded by BEIS capital funding via STFC capital grants ST/K000373/1 and ST/R002363/1 and STFC DiRAC Operations grant ST/R001014/1. DiRAC is part of the National e-Infrastructure. Access to DiRAC resources was granted through Director's Discretionary Time allocations in 2019 and 2020. We acknowledge CINECA for the availability of high performance computing resources and support under the program Accordo Quadro MoU INAF-CINECA ``Nuove frontiere in Astrofisica: HPC e Data Exploration di nuova generazione'' (project ``INA20\_C6A55''). We acknowledge PRACE for awarding us access to SuperMUC-NG at GCS@LRZ, Germany and Marconi at CINECA, Italy. 
The discussion of the results of this work has been facilitated by the Project HPC-EUROPA3 (INFRAIA-2016-1-730897), with the support of the EC Research Innovation Action under the H2020 Programme (grants HPC177WO5I and HPC17MTH1N).
We acknowledge funding by Fondazione Cassa di Risparmio di Firenze under the project ``HIPERCRHEL''.
\\

\bibliographystyle{aasjournal}

\end{document}